\documentclass{article}
\usepackage[dvipdfmx]{graphicx}
\usepackage{geometry}
\usepackage{layout}
\usepackage{amssymb,amsmath}
\usepackage{authblk}
\usepackage{color}
\usepackage[normalem]{ulem}  

\renewcommand\sout{\bgroup \color{red}\ULdepth=-.5ex \ULset}
\newcommand\soutb{\bgroup \color{blue} \ULdepth=-.5ex \ULset}

\def\bal#1\eal{\begin{align}#1\end{align}}
\newcommand{\beq}{\begin{eqnarray}}
\newcommand{\eeq}{\end{eqnarray}}
\newcommand{\p}{\partial}

\newcommand{\vs}[1]{\vspace{#1 mm}}
\newcommand{\hs}[1]{\hspace{#1 mm}}
\newcommand{\bpm}{\begin{pmatrix}}
\newcommand{\epm}{\end{pmatrix}}
\newcommand{\Z}{\mathbb{Z}}

\newcommand{\C}{\mathbb{C}}

\newcommand{\ba}{\left(\begin{array}}
\newcommand{\ea}{\end{array} \right)}

\begin{document}
\title{Field theoretical model of multilayered Josephson junction \\ 
and dynamics of Josephson vortices
}
\author[$1,2$]{Toshiaki Fujimori}
\author[$2,3$]{Hideaki Iida}
\author[$1,2$]{Muneto Nitta}
\affil[$1$]{Department of Physics, Keio University, 
Hiyoshi 4-1-1, Yokohama, Kanagawa 223-8521, Japan}
\affil[$2$]{Research and Education Center for Natural Sciences, Keio University, 
Hiyoshi 4-1-1, Yokohama, Kanagawa 223-8521, Japan}
\affil[$3$]{School of Biomedicine, Far Eastern Federal University, 690950 Vladivostok, Russia}
\maketitle

\thispagestyle{empty}
\begin{abstract}
Multi-layered Josephson junctions are modeled in 
the context of a field theory, and 
dynamics of Josephson vortices trapped 
inside insulators are studied. 
Starting from a theory consisting of complex and real scalar fields 
coupled to a U(1) gauge field 
which admit parallel $N-1$ domain-wall solutions,   
Josephson couplings are introduced weakly between 
the complex scalar fields.
The $N-1$ domain walls behave as insulators separating 
$N$ superconductors,  
where one of the complex scalar fields has a gap. 
We construct the effective Lagrangian on the domain walls, 
which reduces to a coupled sine-Gordon model  
for well-separated walls 
and contains more interactions 
for walls at short distance.
We then 
construct sine-Gordon solitons emerging in an effective theory 
in which we identify Josephson vortices carrying 
singly quantized magnetic fluxes. 
When two neighboring superconductors tend to have 
the same phase, the ground state   
does not change with the positions of domain walls 
(the width of superconductors).
On the other hand, when 
 two neighboring superconductors tend to have 
$\pi$-phase differences, the ground state 
has a phase transition depending on 
the positions of domain walls;
when the two walls are close to each other 
(one superconductor is thin), frustration occurs 
because of the coupling between
the two superconductors besides the thin 
superconductor. 
Focusing on the case of three superconductors separated 
by two insulators, 
we find for the former case that 
the interaction between two Josephson vortices 
on different insulators
changes its nature, i.e., attractive or repulsive, 
depending on the positions of the domain walls. 
In the latter case, 
there emerges fractional Josephson vortices 
when two degenerate ground states appear due to spontaneous charge-symmetry breaking, 
and the number of the Josephson vortices 
varies with the position of the domain walls. 
Our predictions should be verified in multilayered Josephson junctions. 

\end{abstract}

\newpage
\setcounter{page}{1}
\section{Introduction}
 The Josephson effect is one of the most striking macroscopic quantum phenomena, 
which was theoretically predicted and 
experimentally confirmed in the 1960s 
\cite{BDJvar,AR63,S63,R63,DG64}. 
The phenomenon is realized by a system consisting 
of two superconductors 
which are shielded by a thin insulator and are weakly interacted, 
called the Josephson junction. 
Due to the phase difference of macroscopic wavefunctions 
of the two superconductors, 
an electric current is induced even without any voltage difference between the superconductors. 
Now the effect became a basic ingredient in condensed-matter physics 
and is written in many standard textbooks (for example, see Refs.\cite{Tinkham,Barone}). 
Due to recent progress, 
the effect can be seen not only in the standard Josephson junctions 
but also in various weak links of superconductors consisting of new materials 
: graphene \cite{Heersche2007} and topological insulators 
\cite{Veldhorst2012,Williams2012}, for example. 
The Josephson effect is also important for engineering science. 
The superconducting quantum interference device (SQUID) \cite{JLSM} 
and superconducting qubits \cite{DS2013}  
are the typical examples of the application of the phenomenon. 

When a magnetic field is applied parallel to 
a Josephson junction 
made of type-II superconductors,
vortices (magnetic flux tubes)  in 
the type-II superconductors are 
absorbed into the insulator. 
Such magnetic vortices trapped inside an insulator 
are called Josephson vortices or fluxoids 
\cite{Ustinov:1998}. 
The dynamics of Josephson vortices 
can be described by the sine-Gordon model 
\cite{Anderson1964,Barone1971,Rubinstein1970,Scott1973,Witham1975}.  
On the other hand, 
studies of the vortices in various complex setups are frequently done by using 
 the simulations of the Ginzburg-Landau (GL) model:  
3D GL calculation in anisotropic mesoscopic superconductors \cite{Liu2011}, 
vortex-antivortex pair generation in the presence of applied electric current 
\cite{Berdiyorov1}, 
time-dependent calculation of the vortices under an external source 
\cite{Berdiyorov2,Berdiyorov3,Berdiyorov4,Berdiyorov5}, 
and so on. 
Josephson vortex is not a mere conceptual object in theoretical physics, 
but a detectable one:  
it is directly observed by using scanning tunneling microscopy on 
the surface of Si(111)-($\sqrt{7}\times\sqrt{3}$)-In \cite{Yoshizawaetal2014} and 
in a lateral superconductor-normal-superconductor (SNS) network of superconducting Pb nanocrystals linked together by an atomically thin Pb wetting layer \cite{Roditchevetal2015}. 

Some materials have structures similar to Josephson junctions. 
Oxide high-$T_c$ superconductors have a multi-layer structure of superconductors 
(planes of Cu$_2$O) and insulators (other atomic layers) 
\cite{Blatter:1994}.
The coupling between the layered superconductors varies with the materials.  
The coupling in 
BSCCO (Bi$_2$Sr$_2$CaCu$_2$O$_{8+\delta}$) \cite{MTFA88,Setal88_1,Setal88_2,Tetal88} is especially weak, 
and it is known that these materials  
behave like multilayered Josephson junctions, from the analysis of 
their current-voltage characteristics and the specific property of high-frequency electromagnetic waves:
the terahertz laser is produced continuously by BSCCO due to 
its Josephson plasma oscillation 
\cite{Anderson1964,BDJ66,DDFLS68,Oetal07,Welp2013}, 
which is a collective motion of Josephson vortices and superconducting electrons and
 very important for applications in 
engineering \cite{Oetal07,Welp2013}. 
Not only the natural multilayered structure of Josephson junctions, 
but the artificial multilayered superconductors and insulators are also available with the 
development of precise processing technology: 
artificial construction of high-$T_c$ superconductors started already in 
the last 1980s \cite{Triscone1989}, and more recently, 
the experiment of mesoscopic superconducting rings which make a layered 
structure is performed \cite{Bluhm2006}, for example. 
These developments enable us to test the theoretical prediction in various 
experimental setups.

In this study, we propose a simple field theoretical 
model describing multilayered Josephson junctions 
and we study the dynamics of Josephson vortices. 
The model for an $N$-layered Josephson junction 
can be described by 
the $\mathbb{C}P^{N-1}$ model. 
This is a multicomponent extension 
of the previous study of two superconducting layers with one junction  
\cite{Nitta:2012xq,Kobayashi:2013ju}. 
We start from the $U$(1) gauge theory with one real 
and $N$ complex scalar fields,
similarly to the Ginzburg-Landau theory 
for a single superconductor. 
We consider the critical coupling, 
which is known as the Bogomol'nyi-Prasado-Sommerfield 
(BPS) limit in the field theory language. 
This assumption technically simplifies the treatment but is not essential for 
the dynamics of Josephson vortices.  
Taking the strong coupling limit of it, 
we obtain a massive $\mathbb{C}P^{N-1}$ model, 
where $N-1$ parallel domain walls are allowed 
behaving as  insulators.
Then, Josephson terms are introduced 
between superconductors perturbatively.
We construct 
the low-energy effective Lagrangian 
of the domain walls (insulating junctions)   
and find that it reduces to a coupled sine-Gordon 
model  \cite{SBP1993,PS1998} in the limit of well-separated domain walls 
(thick superconductors)  
while it contains more general interaction 
for domain walls at short distances 
(thin superconductors).
The effective theory allows sine-Gordon solitons 
carrying quantized magnetic fluxes,
which we identify as Josephson vortices. 
Focusing on 
the $N=3$ case, 
where two domain walls
(three superconductors and two insulators) 
 exist, 
we investigate 
the dependence of
the effective potential and 
the sine-Gordon solitons 
on the distance of the domain walls 
(the thickness of the middle 
superconductor). 
There are essentially the two cases depending on the Josephson coupling: 
When two neighboring superconductors tend to have 
the same phase, the ground state   
does not depend on the positions of domain walls 
(the width of the middle superconductor).
When two neighboring superconductors tend to have 
the $\pi$-phase differences, the ground state 
has a phase transition depending on the positions of domain walls;
when the two walls are close to each other 
(the middle superconductor is thin), frustration occurs 
because of the coupling between
the two superconductors besides 
the middle superconductor. 
We study dynamics of Josephson vortices 
both as sine-Gordon solitons in the effective theory 
and as full numerical configurations. 
In the unfrustrated case, 
the interaction of Josephson vortices 
at the two neighboring insulators  
changes its nature, i.e., 
the interaction is
attractive when the two insulators are well-separated 
(the middle superconductor is thick), 
while the interaction is repulsive when 
they are close to each other 
(the middle superconductor is thin). 
In the frustrated case, fractional sine-Gordon solitons 
emerge when two degenerate ground states appear due to spontaneous charge-symmetry breaking,   
depending on the distance of two domain walls.  

This paper is organized as follows. 
In Sec.~\ref{sec2}, we construct the field theoretical model of a multilayered Josephson 
junction. We derive the massive 
$\mathbb{C}P^{N-1}$ model 
in the strong coupling limit of 
a $U$(1) gauge theory coupled with real and complex scalar fields,  
 and give the domain wall solutions. 
Then, the Josephson terms are introduced 
to the model and 
the effective theory of domain walls is derived. 
In Sec.~\ref{sec3}, multilayered Josephson junctions in the model are explained. 
After a brief review of the $N=2$ case, where there is one domain wall, 
we study the case of two domain walls in $N=3$. 
In Sec.~\ref{sec4}, we numerically investigate the properties of Josephson vortices:  
the vacuum structure, the interaction between the vortices,  
the profiles of sine-Gordon solitons, energy densities and fluxes 
are studied for various setups. 
Section \ref{sec5} is devoted to a summary and discussion.  
We make a comment on the possibility of realization in 
superconductors and discuss possible extensions  
such as 
supersymmetry and 
non-Abelian (color) superconductors.

\newpage
\section{Field theoretical model of a multi-layered Josephson junction}
\label{sec2}

\subsection{A model without Josephson interactions}
We start with the following $SU(N)$-invariant Abelian-Higgs system\footnote{In this paper, we study only the static problem.
We consider a relativistic theory that shares common 
properties with non-relativistic theories usual for condensed matter 
systems, as far as we concentrate on static problems.
See Refs.~\cite{Kobayashi:2014xua,Kobayashi:2014eqa} for corresponding 
non-relativistic theory. 
} 
\beq
\mathcal L_{A,\phi} ~= - \frac{1}{4 e^2} F_{\mu\nu} F^{\mu\nu} - (D_\mu\phi_a) \overline{(D^\mu\phi_a)} - \frac{\lambda}{4}(|\phi_a|^2 - v^2)^2, \hs{10}
D_\mu &\equiv \partial_\mu - i A_\mu,
\eeq
where $\phi^a$ ($a=1,\cdots, N$) are charged scalar fields, 
$A_\mu$ are Abelian gauge fields and $F_{\mu\nu}$ are  
their field strength. 
In terms of the Ginzburg-Landau model, 
$F_{\mu\nu}F^{\mu\nu}=2(B^2-E^2)$, where $B$ is the magnetic field 
and $E$ the electric field, 
$\phi^a$ are the multicomponent superconductor order parameters,  
the second term correspond to the kinetic term of them and 
their couplings to $E$ and $B$, and 
the third term corresponds to their self-interaction. 
This Lagrangian has the global $SU(N)$ symmetry which rotates the complex scalar fields $\phi_a$. 
The scalar potential is minimized when the scalar fields $\phi_a$ have vacuum expectation values 
(VEV; i.e., condensations) such that 
\beq
|\phi_a|^2 = v^2. 
\eeq
As well as the $U(1)$ gauge symmetry, 
the $SU(N)$ global symmetry is spontaneously broken to $SU(N-1) \times U(1)$
by the nonzero VEVs of the charged scalar fields $\phi_a$. 
Therefore, the low-energy degrees of freedom are 
the Nambu-Goldstone (NG) modes 
parametrizing the complex projective space 
\beq
\C P^{N-1} \simeq \frac{SU(N)}{SU(N-1) \times U(1)}.
\eeq 
In other words, this system is described by the $\C P^{N-1}$ nonlinear sigma model 
when the energy scale is much smaller than 
the masses $ev$ and $\sqrt{\lambda} v$ 
of the massive photon  and massive scaler field, respectively.
[$(ev)^{-1}$ and $(\sqrt{\lambda} v)^{-1}$ are 
the penetration depth and coherence length, respectively.]

\begin{figure}[t]
\begin{center}
\includegraphics[width=140mm]{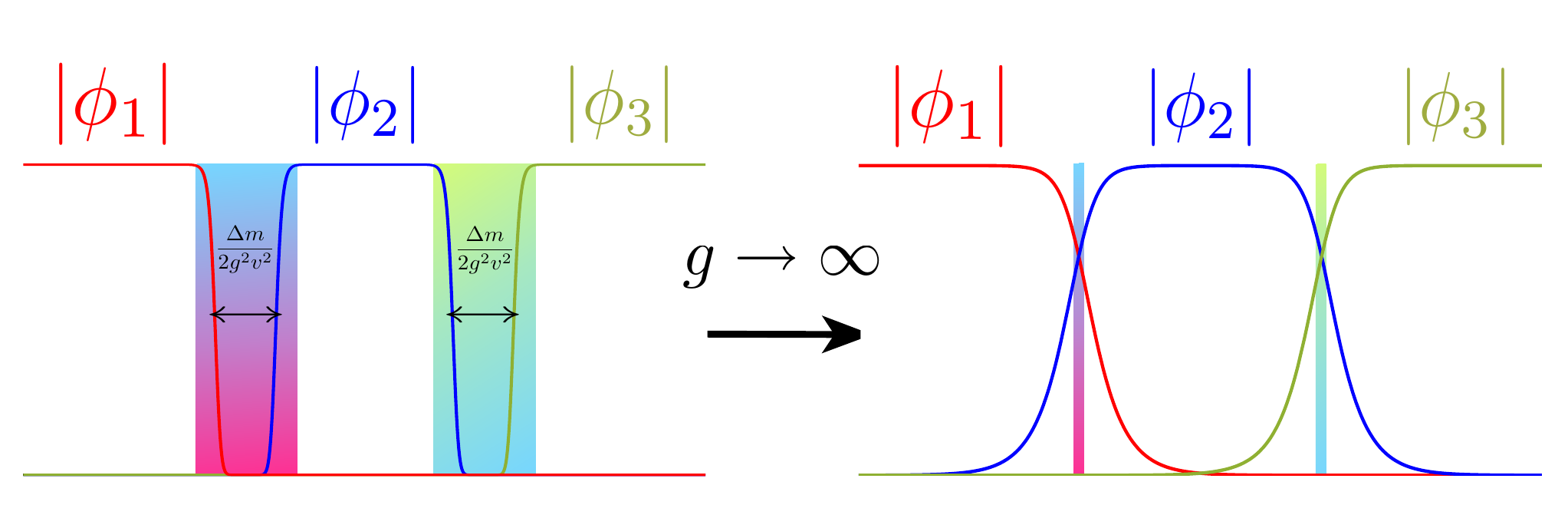}
\caption{Schematic figure of the domain-wall structure of the system for $N=3$ 
for finite $g$ (left) and the limit of infinite $g$ (right, thin-wall limit).  
The regions where finite VEVs (condensations) exist can be viewed as  
bulk superconductors and in between where the domain walls exist  
as insulators of Josephson junctions.}
\label{fig:wall_profile}
\end{center}
\end{figure}

Let us deform the model so that only one of $\phi_a$ can have a nonzero VEV 
by introducing a neutral real scalar field $\Sigma$ and 
mass parameters $m_a~(a=1,\cdots,N)$. 
We then add the following terms into the original Lagrangian: 
\beq
\mathcal L_{\Sigma} = -\frac{1}{g^2} \partial_\mu \Sigma \, \partial^\mu \Sigma - \sum_{a=1}^N (m_a - \Sigma)^2|\phi_a|^2. 
\eeq
The potential in the total Lagrangian $\mathcal L \equiv \mathcal L_{A,\phi} + \mathcal L_{\Sigma}$ 
is minimized when the scalar fields satisfy
\bal
\sum_{a=1}^N (m_a-\Sigma)^2|\phi_a|^2=0,
\hs{10}
\sum_{a=1}^N |\phi_a|^2=v^2.
\eal
There are $N$ degenerated ground states 
labeled by $b\in\{1,2,\cdots,N\}$, 
each of which is characterized by the following VEVs of the scalar fields
\bal
\phi_a=
\left\{
\begin{array}{cc}
v & \mbox{for \ } a=b \\
0 & \mbox{for \ } a \neq b
\end{array}
\right.,
\hs{10} 
\Sigma = m_b.
\label{eq:VEV}
\eal
In the following, we consider domain walls interpolating these discrete generate vacua 
(see Fig.\,\ref{fig:wall_profile}). 
It will turn out that they have the role of insulating junctions in the following discussions. 

To discuss the property of the domain walls, 
it is convenient to consider a limit in which the system is described by a simplified model. 
Since the mass of the fluctuation of $\Sigma$ is $g v$, 
its dynamics is decoupled in the low-energy limit $E \ll g v$. 
The parameters $m_a~(a=1,\cdots,N)$ explicitly break 
the $SU(N)$ symmetry to the Cartan subgroup $U(1)^{N-1}$
and hence give masses to the ${\C} P^{N-1}$ NG modes.
Therefore, in the low-energy regime where $E \approx m_a \ll ev,\, \sqrt{\lambda} v,\, gv$, 
the system is described by the $\mathbb{C}P^{N-1}$ nonlinear sigma model with mass terms. 
The corresponding action can be obtained by taking the limit $e,\, g,\, \lambda\rightarrow \infty$, 
and then eliminating the heavy degrees of freedom $A_\mu$ and $\Sigma$ 
by solving their equations of motion:  
\bal
A_\mu = \frac{i}{2v^2} ( \bar \phi_a \partial_\mu\phi_a - \phi_a \partial_\mu \bar \phi_a ),
\hs{10} 
\Sigma = \frac{1}{v^2}\sum_{a=1}^N m_a |\phi_a|^2,  
\label{eq:sol_infinite_g}
\eal 
where the charged scalar fields must satisfy the constraint $|\phi_a|^2 = v^2$.
To write down the effective Lagrangian, 
it is convenient to introduce the inhomogeneous coordinates of $\mathbb{C}P^{N-1}$ 
defined by
\bal
(\phi_1,\cdots,\phi_{N-1},\phi_N)
=
\frac{v}{\sqrt{1+|u_i|^2}}(u_1,\cdots,u_{N-1},1), 
\label{eq:coordinates}
\eal
where we have used the $U(1)$ gauge transformation to fix the overall phase so that $\arg \phi_N = 0$. 
Then, we can rewrite the original Lagrangian in the limit $e,g,\lambda \rightarrow \infty$ into the following form
\bal
{\cal L}=-v^2g_{i\bar j}(\partial_\mu u^i\partial^\mu\bar u^j + \Delta_i\Delta_j u^i\bar u^j),
\label{eq:L_CP}
\eal
where $\Delta_i\equiv m_i-m_N$ and 
the Fubini-Study metric of $\C P^{N-1}$ is given by
\bal
g_{i\bar j}=\frac{\partial^2}{\partial u^i\partial \bar u^j}
\ln(1+|u_k|^2).
\eal

Figure\,\ref{fig:wall_profile} shows a domain-wall configuration for $N=3$ before and after taking the limit. 
The regions with different condensations are separated by the domain walls,  
whereas there is no condensation inside the walls. 
Thus, the domain walls can be regarded as the insulating junctions separating 
each superconductor in terms of Josephson junctions in condensed matter physics. 
 The width of the wall can be estimated as $\Delta m/2g^2v^2$ (for $\lambda \approx g^2$) 
and hence the sigma model limit corresponds to a thin-wall limit. 

\subsection{Domain wall solutions: inserting insulators}
Next, let us consider domain-wall solutions 
in the massive $\C P^{N-1}$ nonlinear sigma model in 
Eq.\,\eqref{eq:L_CP} 
\cite{Gauntlett:2000ib}.
Suppose that the fields $u_i$ depend only on one of the spatial coordinates $x$.  
Then the energy of the system can be rewritten as
\bal
E = v^2 \int d^d x \left[
g_{i\bar j}(\partial_x u^i-\Delta_i u^i)
\overline{(\partial_x u^j-\Delta_j u^j)}
+\partial_x \left(\frac{\Delta_i |u_i|^2}{1+|u_k|^2}\right)
\right].
\eal
Since the total derivative term is a constant for a fixed boundary condition, 
the energy is minimized when $u_i$ satisfy 
\bal
\partial_x u^i = \Delta_i u^i.
\eal
The domain-wall solution is given by
\bal
u^i = \exp \left( \Delta_i x + \xi_i + i \theta_i \right),
\label{eq:wall_sol}
\eal
where $\xi_i$ and $\theta_i$ are arbitrary parameters.
Going back to the original description in terms of $\phi_a$, 
we can see that there are domain walls interpolating the regions with different condensations
(see Fig.\,\ref{fig:wall_profile}). 

The parameters $\xi_i$ are related to the domain-wall positions, 
which can be read from the energy density of the configuration
\beq
\mathcal E ~=~ \frac{v^2}{2} \p_x^2 \ln ( 1 + |u_i|^2 ) ~=~ 
\frac{v^2}{2} \p_x^2 \log \left( 1 + \sum_{i=1}^{N-1} e^{2(\Delta_i x + \xi_i)} \right). 
\eeq
This is small in the regions where only one of the terms in the logarithm is large. 
Therefore, the domain walls are localized where any two terms in the logarithm  
are of the same order of magnitude. 
Suppose that the masses are ordered as $m_1<m_2<\cdots<m_N$.
Then we can determine their positions as \cite{Isozumi:2004jc}
\bal
X_i=-\frac{\xi_i-\xi_{i+1}}{\Delta_i-\Delta_{i+1}} \ \ (i=1,\cdots,N-1),
\eal
where $\xi_N=0$ and $\Delta_N=0$. 

The parameters $\theta_i~(i=1,\cdots,N-1)$ are related to the phases of $\phi_a$.
In the regions where only one of $\phi_a$ is nonzero, 
$\arg \phi_a$ can be eliminated by gauge transformations.
On the $i$th domain wall ($x=X_i$), 
there is an overlap of $\phi_i$ and $\phi_{i+1}$, 
so that the relative phase $\arg \phi_i - \arg \phi_{i+1}$ cannot be eliminated. 
Since those relative phases are rotated by the $U(1)^{N-1}$ global symmetry, 
$\theta_i$ can be viewed as the NG modes associated with 
the $U(1)^{N-1}$ symmetry broken by the domain walls. 

Now let us derive the low-energy effective model describing the dynamics of 
the degrees of freedom living on the domain walls. 
To write down the effective Lagrangian, 
it is convenient to introduce the complex moduli parameters $\varphi_i~(i=1,\cdots,N-1)$ defined by
\bal
\varphi_i=\xi_i+i\theta_i.
\eal
In the low-energy regime, 
we can assume that $\varphi_i$ weakly depends on time and the coordinates on the domain walls. 
Then substituting the solution $u^i=e^{\Delta_i x+\varphi_i}$ into the original Lagrangian \eqref{eq:L_CP}, 
we obtain the effective action of the form 
\bal
{\cal L}_{\rm eff} = - \sum_{i=1}^{N-1} T_i + {\cal G}_{i\bar j} \left( \frac{\p \varphi^i}{\p t} \overline{\frac{\p \varphi^j}{\p t}} - \frac{\p \varphi^i}{\p y^\alpha} \overline{\frac{\p \varphi^j}{\p y^\alpha}} \right), 
\eal
where $y^\alpha$ are the spatial directions perpendicular to $x$. 
The constants $T_i = v^2 (m_{i+1}-m_i)$ are the tensions (energy per unit area) of the domain walls and 
the moduli space metric ${\cal G}_{i\bar j}$ is given by 
\bal
{\cal G}_{i\bar j} = v^2\int dx \, \frac{\partial^2}{\partial\varphi^i \partial\bar\varphi^j}
\ln\left( 1 + \sum_{k=1}^{N-1} e^{2 \Delta_k x + \varphi^k + \bar \varphi^k} \right) .
\eal
The effective theory in the nonrelativistic case can be found 
in Ref.~\cite{Kobayashi:2014xua} for the ${\mathbb C}P^1$ model.

In the following, 
we consider modifications to the domain-wall effective Lagrangian 
in the presence of Josephson terms 
and then study the solitons, 
assuming that the domain walls are static and the Josephson terms are weak. 

\subsection{Introduction of Josephson interactions}
In this subsection, 
we construct the effective Lagrangian of the domain walls in the presence of Josephson terms.  
Let us consider the following deformation term which breaks the $U(1)^{N-1}$ global symmetry: 
\bal 
V_{\rm J} = \sum_{(a,b)} \beta_{ab} \, \bar{\phi}^{a} \phi^{b}, 
\label{Josephsonterms}
\eal
where $\beta_{ab}$ is a Hermitian matrix whose diagonal elements are all equal to zero. 
This term induces a potential term in the domain-wall effective action. 
For small $\beta_{ab}$, the leading order effective potential can be obtained 
by simply substituting the domain-wall solution Eq.\,\eqref{eq:wall_sol} into $V_{\rm J}$ and integrating over $x$:
\beq
V^{\rm eff}_{\rm J} = \sum_{a < b} \gamma_{ab}(\xi) \, \cos( \theta_a - \theta_b + \arg \beta_{ab}), \hs{5}
\gamma_{ab}(\xi) = 2 |\beta_{ab}| \int dx \, \frac{e^{(\Delta_a + \Delta_b) x + \xi_a + \xi_b}}{\sum_{a=1}^{N} e^{2(\Delta_a x+\xi_a)}},
\eeq
where $\Delta_N = \xi_N = \theta_N = 0$.

The domain-wall effective action with this potential can 
be regarded as multilayered Josephson junctions; 
when the expectation values of $\phi_i$ are localized in different domains, 
they can be viewed as bulk superconductors,   
and the domain walls in between correspond to thin insulators  as depicted in Fig.\,\ref{fig:wall_profile}.  
Since the essence of the model as the multilayered Josephson junctions 
is summarized in the $N=3$ case, 
we focus on concrete calculations in the $N=3$ case
after reviewing the case of $N=2$.

\section{Multi-layered Josephson junctions}
\label{sec3}
\subsection{A single Josephson junction of two superconductors 
and a Josephson vortex in it: a review}

For $N=2$ with $-m_1=m_2=m/2$, 
the domain wall solution in the massive $\C P^1$ model takes the form
\cite{Abraham:1992vb}
\bal
u = e^{m(x-X) + i \theta},
\eal
where $X$ and $\theta$ are arbitrary parameters corresponding to the position and phase of the domain wall.  
In the case of $N=2$, we can always redefine the phase of $\phi^a$ 
so that the coupling constant $\beta$ in the Josephson term 
$V_{\rm J} = \beta \, \bar \phi^1\phi^2 + {\rm c.c.}$ is real and positive. 
For small $\beta$, 
the effective action on the domain wall is given by the sine-Gordon model \cite{Nitta:2012xq}
\bal
{\cal L}_{\rm eff} = - m v^2 - \frac{v^2}{2m} \Big[ m^2 (\partial_\mu X)^2 + (\partial_\mu \theta)^2
+ 2\pi \beta \cos\theta \Big].
\eal 
The potential term has the minimum at $\theta=\pi~({\rm mod} \ 2\pi)$. 

As is well known, the sine-Gordon model has kink solutions 
which are characterized by a nontrivial winding of $\theta$. 
The equation describing static kink solutions can be found 
by setting $X=const.$, assuming that $\theta$ depends only on a spatial coordinate $y$ 
and rewriting the energy density as
\beq
\mathcal E_{\rm eff} = mv^2 \left( 1 - \frac{\pi \beta}{m^2} \right) +  
\frac{v^2}{2m} \left[ \left(\p_y \theta \pm 2 \sqrt{\pi \beta} \cos \frac{\theta}{2} \right)^2 
\mp 8 \sqrt{\pi \beta} \, \p_y \sin \frac{\theta}{2} \right].
\label{eq:SG_comp}
\eeq
This is minimized when $\theta$ satisfies
\bal
\p_y \theta \pm 2 \sqrt{\pi \beta} \cos \frac{\theta}{2} = 0. 
\eal 
and the solution is given by
\bal
\theta^{\pm}(y) = 4 \arctan \exp\left[ \pm \sqrt{\pi\beta} (y-Y) \right] + \pi,
\eal
where $Y$ is an arbitrary parameter corresponding to the kink position. 
The total derivative term in Eq.\,\eqref{eq:SG_comp} gives the mass of the kink:
\beq
M_{\rm kink} ~=~ \frac{4 \sqrt{\pi \beta} v^2}{m} \int dy \, \p_y \sin \frac{\theta}{2} ~=~ \frac{8 \sqrt{\pi \beta} v^2}{m}.
\eeq
This object has a quantized magnetic flux:  
using Eq.\,\eqref{eq:sol_infinite_g}, we find that
\beq
\int dx dy \,  F_{xy}  ~=~ - \int dx dy \, \frac{m}{2 \cosh^2 m(x-X)} \p_y \theta^\pm ~=~ \mp 2 \pi.
\eeq
This is precisely a Josephson vortex, 
which is a magnetic vortex trapped inside an insulator 
\cite{Ustinov:1998}.

\begin{figure}[t]
\begin{center}
\includegraphics[width=80mm]{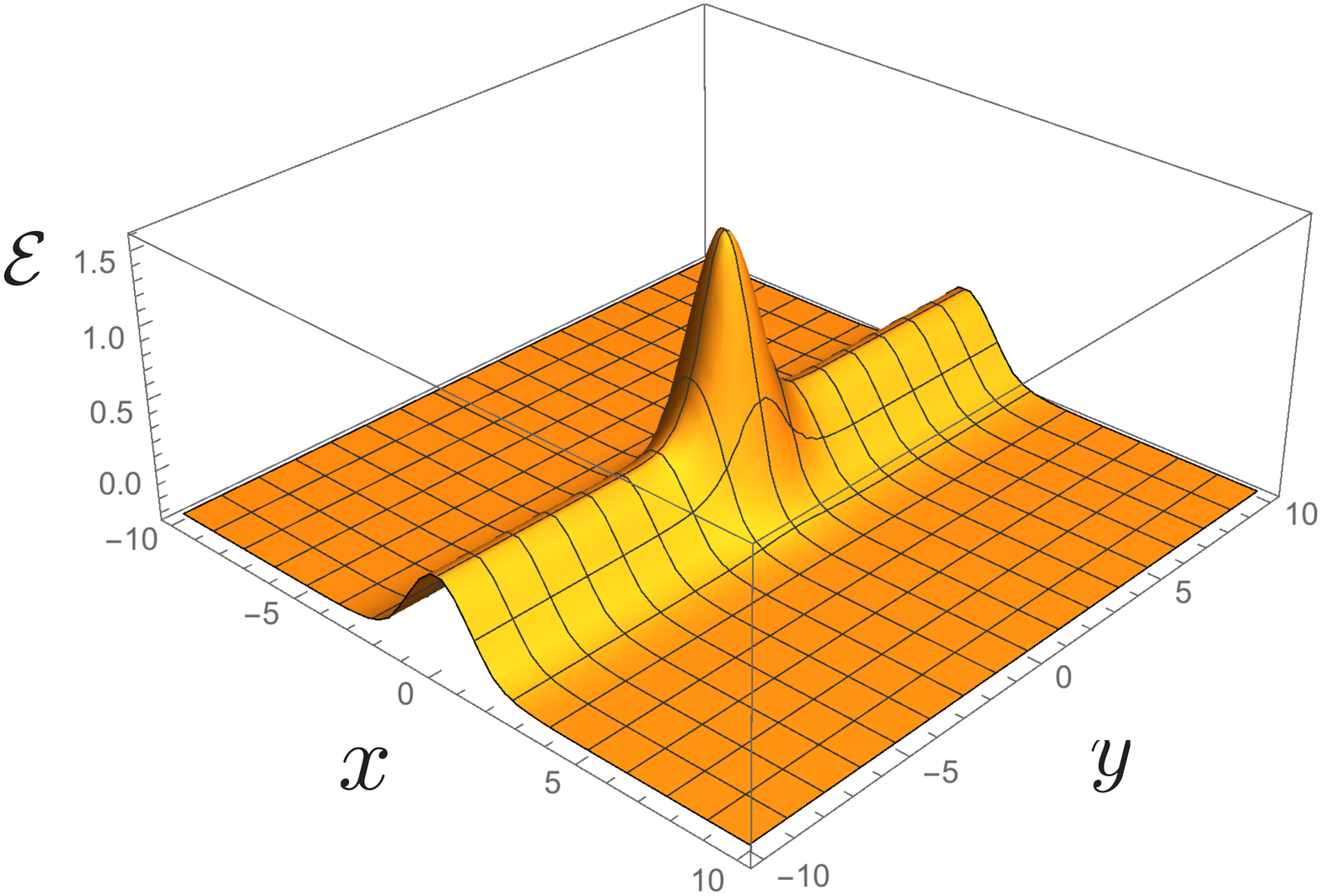}
\caption{Energy density profile of a numerical solution for $\lambda=e=g=v=-m_1=m_2=\beta=1$.}
\label{fig:energydensity}
\end{center}
\end{figure}

Figure\,\ref{fig:energydensity} shows a numerical solution of 
the original model without taking the sigma model limit. 
The domain wall is localized along the line $x=0$, 
on which the kink is localized at $y=0$. 

In the case of $N=2$,  
the phase of $\beta$ can always be absorbed into a constant shift of $\theta$, 
i.e., a redefinition of the phases of $\phi^a$. 
As we will see, in the case of $N=3$, 
one of $\arg \beta_{ab}$ cannot be absorbed by shift of $\theta_a$ and 
the property of the kinks depends on its value.

\subsection{Three-layered Josephson junctions}  
Next, let us consider the $N=3$ case. 
In the following, we set
\bal
(m_1, m_2, m_3)=(-m,0,m),
\eal
for simplicity. 
By the redefinition $\phi_a \rightarrow e^{i \alpha_a} \phi_a$, 
the phases of $\beta_{ab}$ are shifted as
\beq
\beta_{ab} \rightarrow \beta_{ab} e^{i(\alpha_a - \alpha_b)}. 
\eeq
This implies that the phase of $\beta_{12} \beta_{23} \beta_{31}$ does not change, 
and hence there is a physical phase parameter 
which cannot be eliminated by the redefinition. 
By appropriately choosing the phases of $\phi_a$, 
we can always set 
\beq
\arg \beta_{12} = \arg \beta_{23} = \arg \beta_{31} \equiv \vartheta.
\eeq
When $\vartheta = 0$ or $\vartheta=\pi$, 
the Hermitian matrix $\beta_{ab}$ becomes a real symmetric matrix, 
so that the Josephson term preserves the charge conjugation symmetry
\beq
\phi_a \rightarrow \bar \phi_a. 
\eeq
In the following, we focus on these two special cases: 
$\beta_{ab}$ are all positive ($\vartheta = 0$) or negative ($\vartheta = \pi$). 

To write down the effective action of the domain walls, 
it is convenient to use the phase differences 
\bal
\theta_{12} \equiv \theta_1-\theta_2, \hs{10} 
\theta_{23} \equiv \theta_2.
\eal
Note that $\theta_a = \arg \phi_a$ and we have set $\theta_3 = 0$ in Eq.\,\eqref{eq:coordinates} by using the gauge transformation. 

\begin{figure}[t]
\begin{center}
\begin{tabular}{ccc}
\includegraphics[width=50mm]{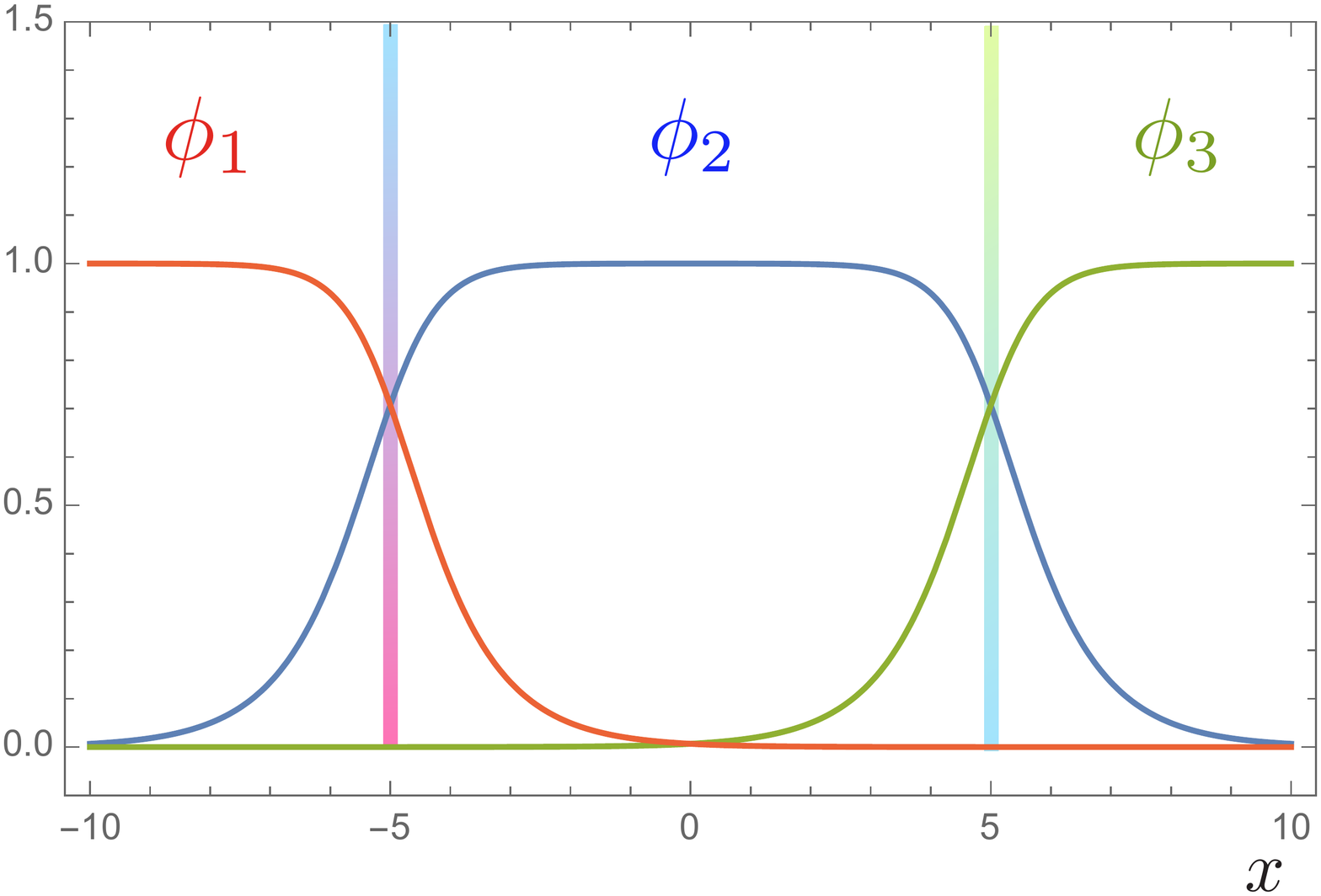} &
\includegraphics[width=50mm]{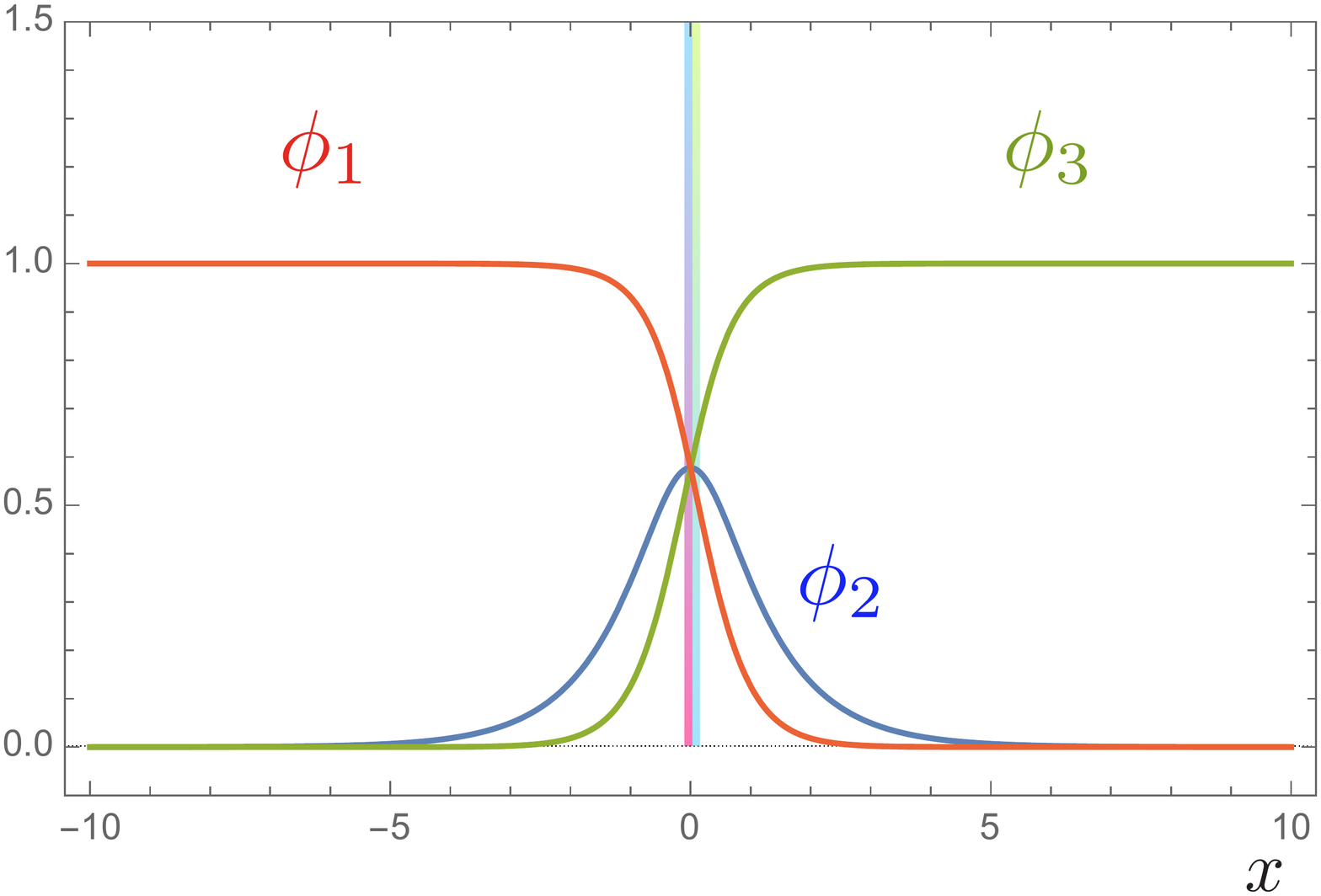} &
\includegraphics[width=50mm]{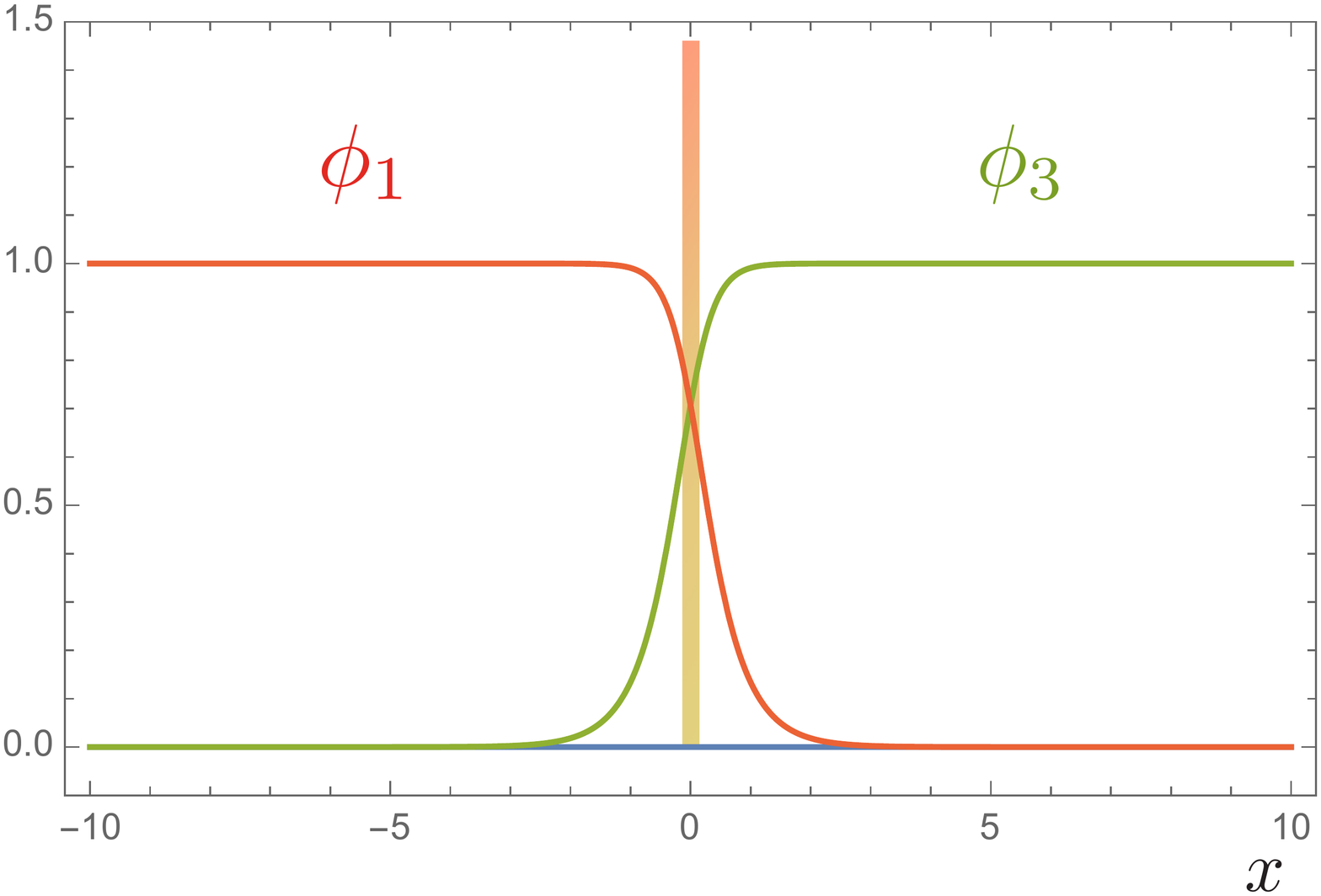} \\
(a) $-X_1=X_2=5$ & (b) $-X_1=X_2=0$ & (c) $-X_1 = X_2 = - 10$
\end{tabular}
\caption{Domain-wall configurations with $m=1$. 
When $X = X_2 - X_1 > 0$, 
$X_1$ and $X_2$ can be viewed as the positions of the walls. 
As $X$ becomes smaller, the walls approach each other and 
they are almost overlapping around $X \approx 0$. 
For $X < 0$, the condensation of $\phi_2$ starts to decrease and disappears in the $X \rightarrow - \infty$ limit. }
\label{fig:2-wall}
\end{center}
\end{figure}

In this setup, the domain-wall solution is given by
\bal
u_1 &= e^{-2m \left( x-\frac{X_1+X_2}{2} \right) + i (\theta_{12} + \theta_{23})} , \nonumber\\
u_2 &= e^{-m(x-X_2) + i \theta_{23}},
\eal
where $X_1$ and $X_2$ are the positions of two domain walls (see Fig.\,\ref{fig:2-wall}). 

\begin{figure}[t]
\begin{minipage}{0.5\hsize}
\begin{center}
\includegraphics[width=65mm]{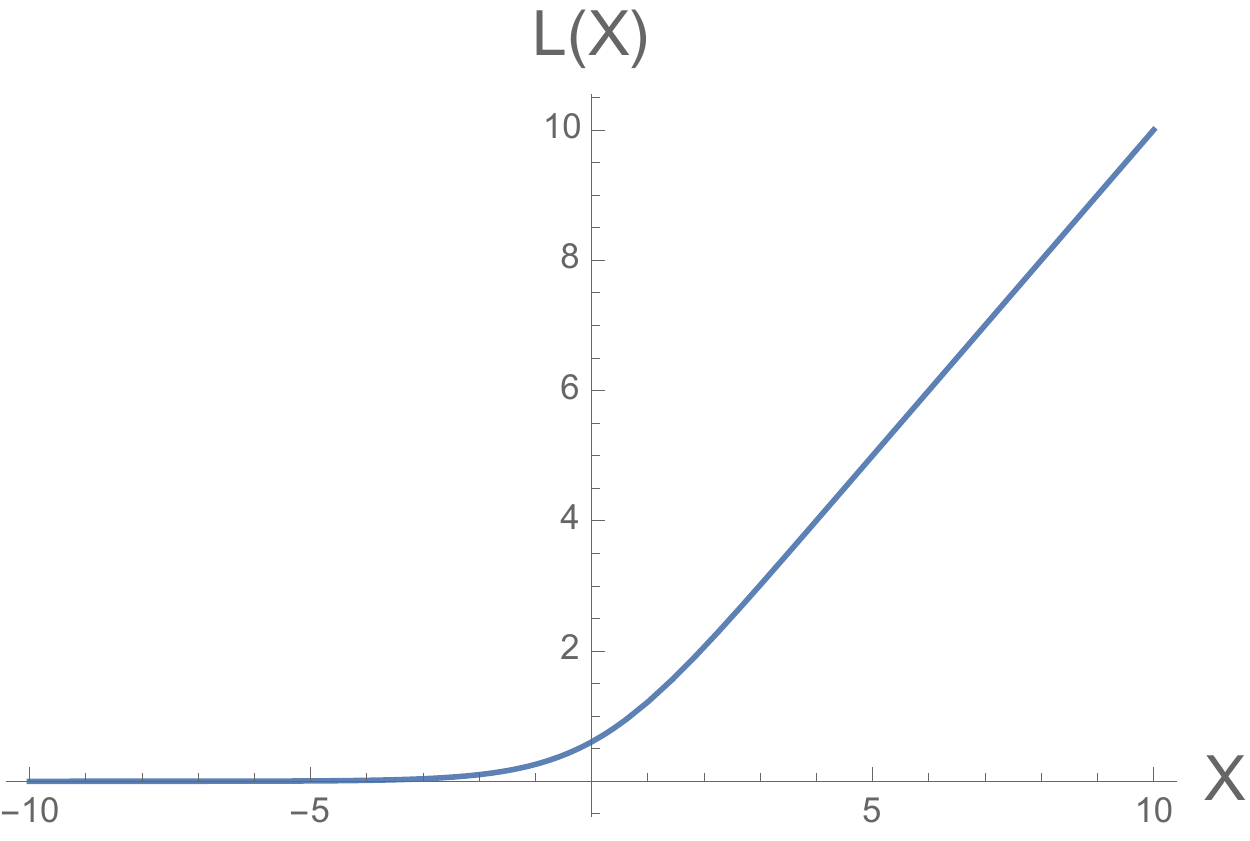}
\caption{The relative distance $L(X)$ against $X$.}
\label{fig:relative_distance}
\end{center}
\end{minipage}
\begin{minipage}{0.5\hsize}
\begin{center}
\includegraphics[width=8cm]{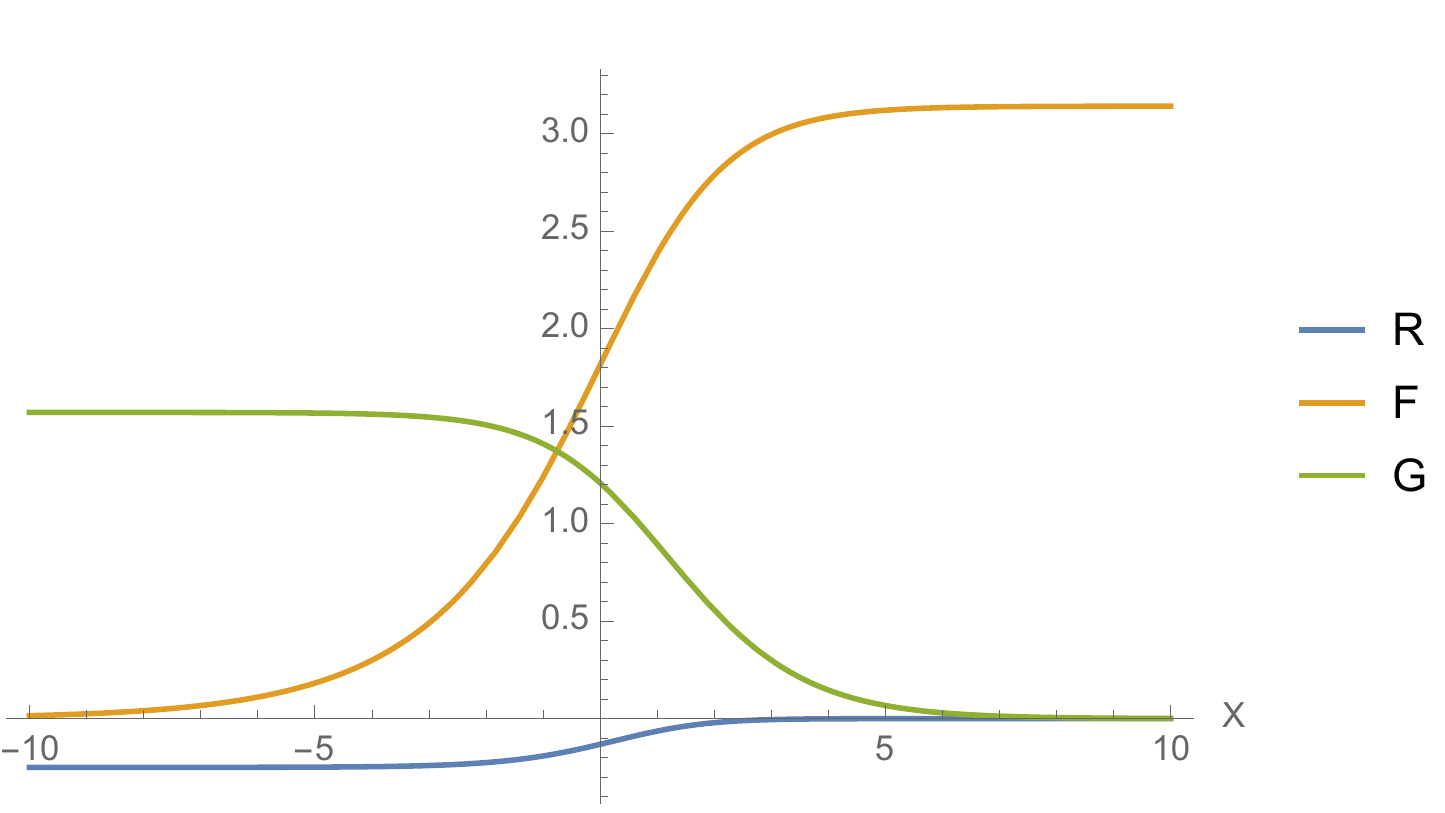}
\caption{$R(X)$, $F(X)$ and $G(X)$ against $X$.}
\label{fig:RFG}
\end{center}
\end{minipage}
\end{figure}

Here we consider the large-tension limit $(m^2 \gg \beta)$ in which dynamics of $X_1$ and $X_2$ 
are negligible
\footnote{In addition, we need the condition $\exp(-mL)\gg \beta/m^2$, 
to satisfy the potential from the 2nd-order perturbation induced 
by the interaction of 1st and 2nd layer and that of 2nd and 3rd layer, 
which is of order $\beta^2/m^2$, is negligible compared to 
the 1st-order interaction of 1st and 3rd layer, which is of order  
$\beta\exp(-mL)$.}. 
The phase part of the effective Lagrangian takes the form
\bal
{\cal L}_{\rm eff}~
&= \
\frac{v^2}{2m}\Big[ (\partial_y \theta_{12})^2+(\partial_y \theta_{23})^2 \Big]
+ \frac{v^2}{m} R(X) \, (\partial_y\theta_{12}-\partial_y\theta_{23})^2
+V_{\rm eff}, \phantom{\Bigg)} \label{eq:Leff} \\
R(X) &\equiv \ \frac{1}{e^{2mX}-4} \big[ 1- m L(X) \big], \phantom{\Bigg]} \nonumber
\eal
where $X$ and $L(X)$ are given by
\beq
X ~\equiv~ X_2-X_1, \hs{10}
L(X) ~\equiv~ \frac{1}{m} \frac{e^{mX}}{\sqrt{e^{2mX}-4}} {\rm cosh}^{-1}(e^{mX}/2).
\eeq
The function $L(X)$ can be viewed as the relative distance between the walls 
(see Appendix D of \cite{Eto:2008mf} for more details).
Since $L(X) \approx X$ for large $X$, the parameter $X$ can be viewed as 
the asymptotic relative distance as we have seen in the previous section. 
For negative $X$, the function $L(X)$ gives the precise definition of the relative distance (see Fig.\,\ref{fig:relative_distance}).

The effective potential $V_{\rm eff}$ is given by
\bal
V_{\rm eff} ~
&= ~ F(X)(\beta_{12} \cos\theta_{12} + \beta_{23} \cos\theta_{23})
 + G(X)\beta_{13} \cos(\theta_{12} + \theta_{23}), \phantom{\Bigg)} \\
& F(X) \equiv \frac{\pi}{\sqrt{1+2e^{-mX}}}, \ \hs{8}
G(X) \equiv  2 m e^{-m X} L(X). \nonumber
\label{eq:29}
\eal
Since the potential depends only on $mX$, 
we set $m=1$ in the following.  
By the redefinition of the coupling constants $\beta_{ab} \rightarrow v^2 \beta_{ab}$, 
the parameter $v^2$ becomes an overall constant of the effective action, 
so that we can set $v^2=1$ in the classical discussion.  

Figure\,\ref{fig:RFG} shows the plot of the functions $F(X)$, $G(X)$ and $R(X)$. 
Since $G(X)$ and $R(X)$ are small for large $X$,  
the interaction between $\theta_{12}$ and $\theta_{23}$ ($\arg \phi_1$ and $\arg \phi_3$) 
is negligible for well separated walls 
and hence the effective action reduces to that for two independent walls. 
On the other hand, the interaction terms become relevant when two walls approach each other.  
Note that, for in the limit $X \rightarrow - \infty$, 
the effective Lagrangian is independent of $\theta_{12}-\theta_{23}$
 and reduces to that of a single wall depending only on $\theta_{12}+\theta_{23} = \arg\phi_1 - \arg\phi_3$. 

The structure of minima of $V_{\rm eff}$ depends on the signs of $\beta_{ab}$. 
In the following, we restrict ourselves to the case where $\beta_{12}=\beta_{23}=\beta_{31} \equiv \beta$. 
Then, there are two cases: $\beta>0$ and $\beta<0$. 
We study the potential, the vacuum structure, 
and the properties of sine-Gordon solitons in each case.

To find solutions of equations of motion, 
we numerically solve the gradient flow equations
\beq
\frac{\p \theta_{12}}{\p t} = - \frac{\delta S_{\rm eff}}{\delta \theta_{12}}, \hs{10}
\frac{\p \theta_{23}}{\p t} = - \frac{\delta S_{\rm eff}}{\delta \theta_{23}}, 
\eeq
where $t$ is a fictitious time and $S_{\rm eff}$ is the effective action corresponding to Eq.\,\eqref{eq:Leff}. 
Starting with an initial condition with a nontrivial topological number, 
we can find a minimum of $S_{\rm eff}$ by taking the $t \rightarrow \infty$ limit. 
Although we cannot take the $t \rightarrow \infty$ limit if there is no stable minimum in the topological sector, 
quasistable configurations can be obtained by solving the gradient flow equation 
for a sufficiently long time interval. 


\section{Interaction between Josephson vortices}
\label{sec4}

\subsection{$\beta <0$: the same phases}

Here, we study the properties of sine-Gordon solitons in the effective theory 
with $\beta_{12}=\beta_{23}=\beta_{31} \equiv \beta <0$.

\subsubsection{Ground state}  
In this case, the minimum of $V_{\rm eff}$ is always located at 
$\theta_{12}=\theta_{23}=0$ (mod $2\pi$) irrespective of the distance between two walls.

\begin{figure}[t]
\begin{center}
\includegraphics[width=5cm]{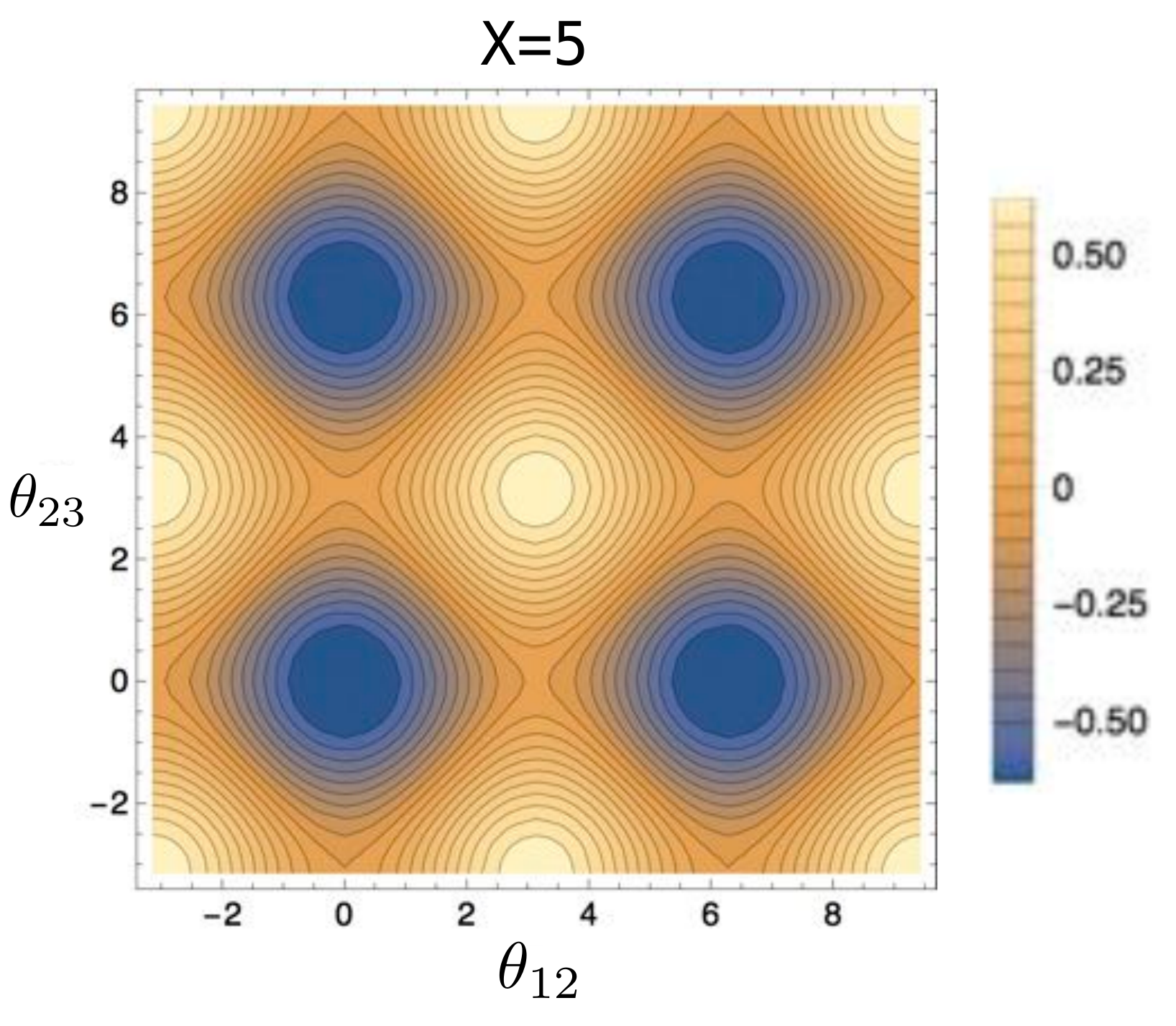}
\hs{10}
\includegraphics[width=5cm]{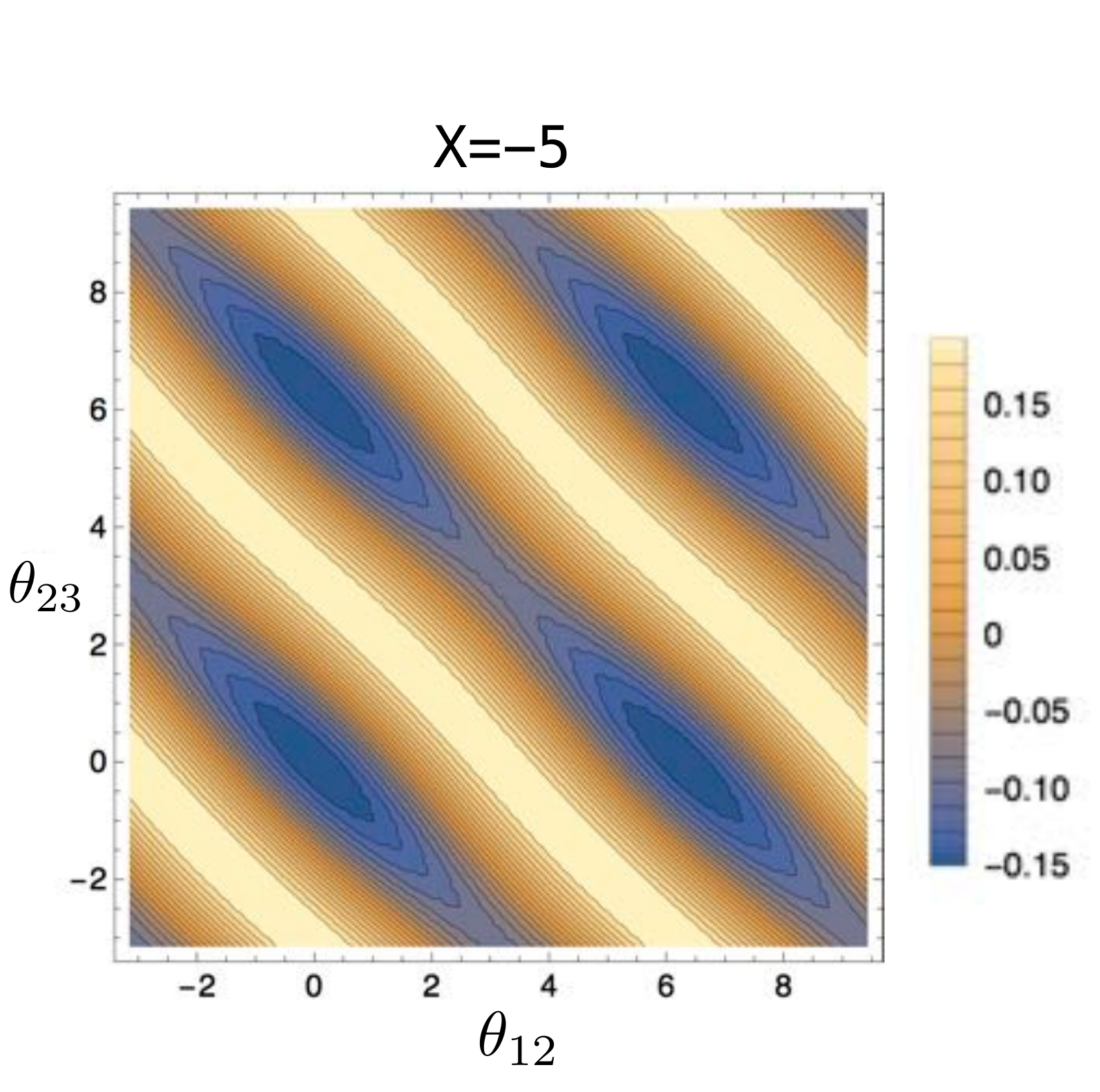}
\caption{The potential $V_{\rm eff}$ for $X=5$ and $-5$ 
with $\beta_{12}=\beta_{23}=\beta_{31}=-1/10$. 
Note that the points $(\theta_{12} + 2 \pi n , \theta_{23}+ 2 \pi m)$ with any $n, m \in \Z$ are 
identified with $(\theta_{12}, \theta_{23})$.}
\label{fig:effective_potential}
\end{center}
\end{figure}

Figure\,\ref{fig:effective_potential} shows effective potentials 
$V_{\rm eff}$ at $X=5,\,-5$ and $\beta=-1/10$. 
Although the shapes of the potential are different, 
the minima of the potential can be seen at $\theta_{12}=\theta_{23}=0$ (mod $2\pi$) in both cases.  

\subsubsection{$(1,1)$: the vortex-vortex interaction}  

\begin{figure}[h]
\begin{center}
\includegraphics[width=17cm]{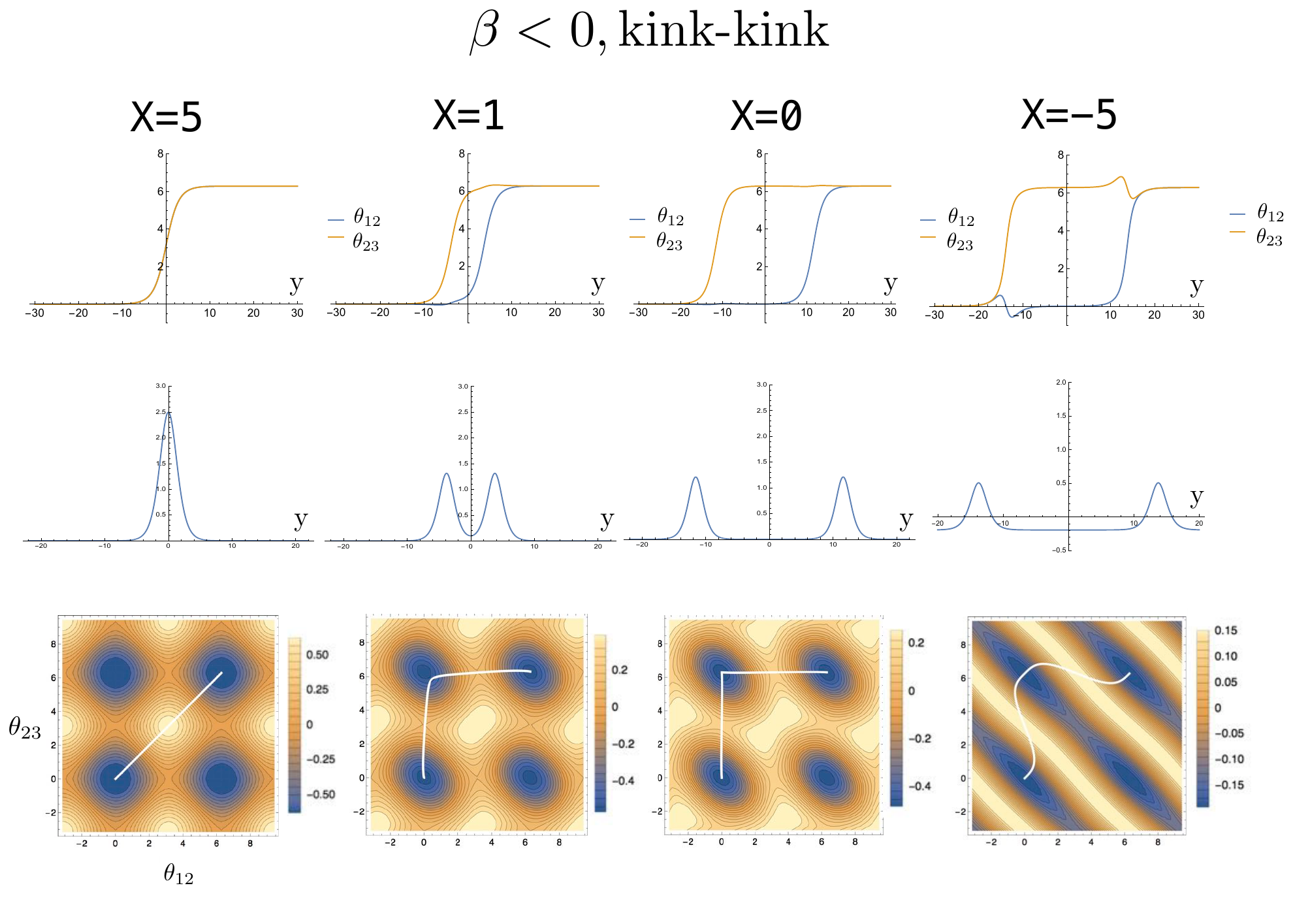}
\caption{$(1,1)$ configurations with $\beta_{12}=\beta_{23}=\beta_{31}=-1/10$. 
The distance between two domain walls is denoted in the upper part of the figure ($X=5,1,0,-5$).
The upper panels are the profiles of $\theta_{12}$ and $\theta_{23}$ against $y$, 
the middle panels are energy densities, 
and lower panels are the contour plot of the effective potential $V_{\rm eff}$ and 
soliton profiles in $\theta_{12}$-$\theta_{23}$ plane. It is clearly seen that the localized energy density around $y=0$ at $X=5$ splits into two peaks and they repel each other at $X=1, 0,$ and $-5$. 
Note that only $\theta_{12}+\theta_{23}$ is physical in the small-$X$ limit.}
\label{fig:negative_(1,1)}
\end{center}
\end{figure}

\begin{figure}[t]
\begin{center}
\includegraphics[width=60mm]{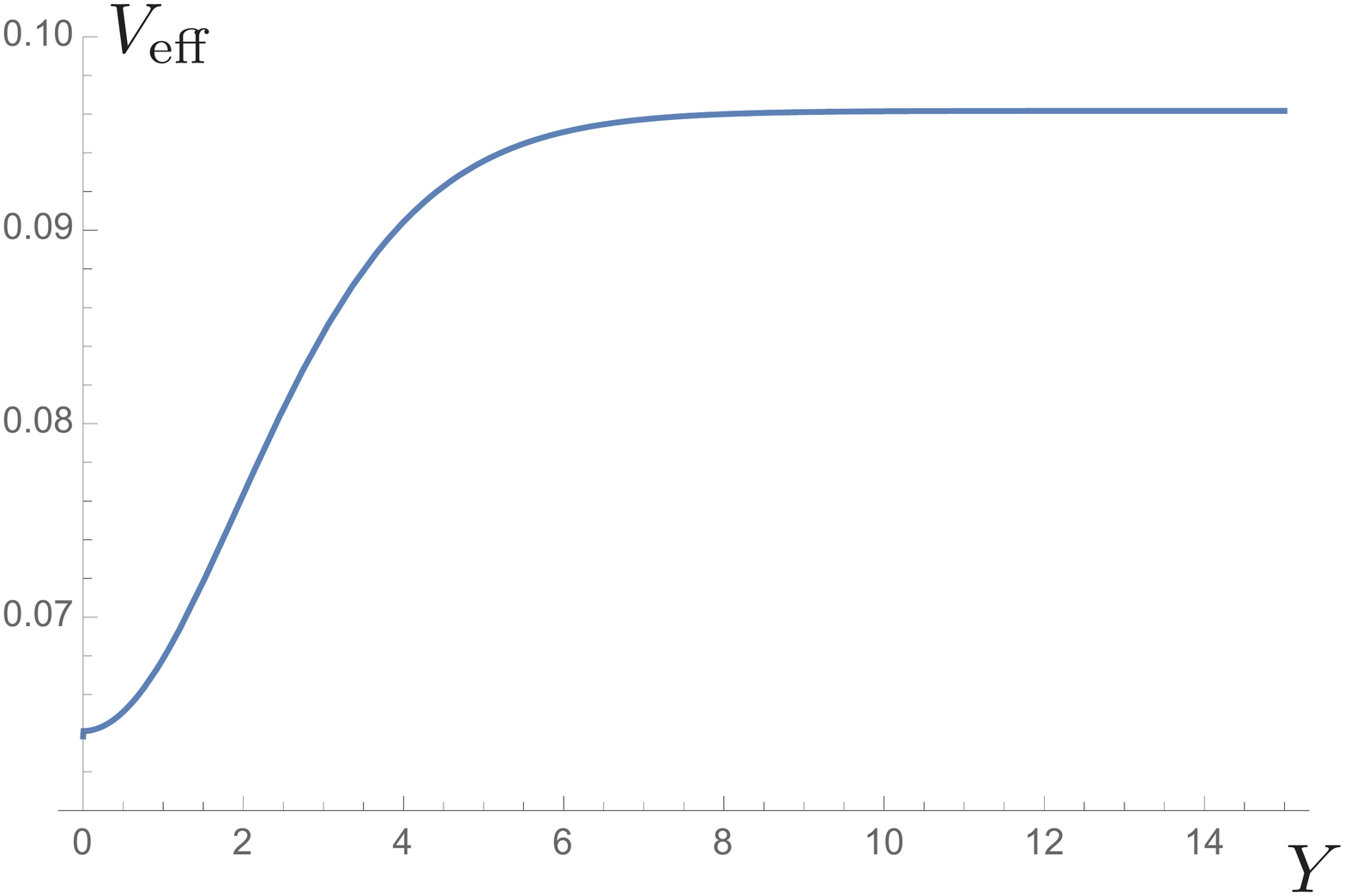} \hs{10}
\includegraphics[width=60mm]{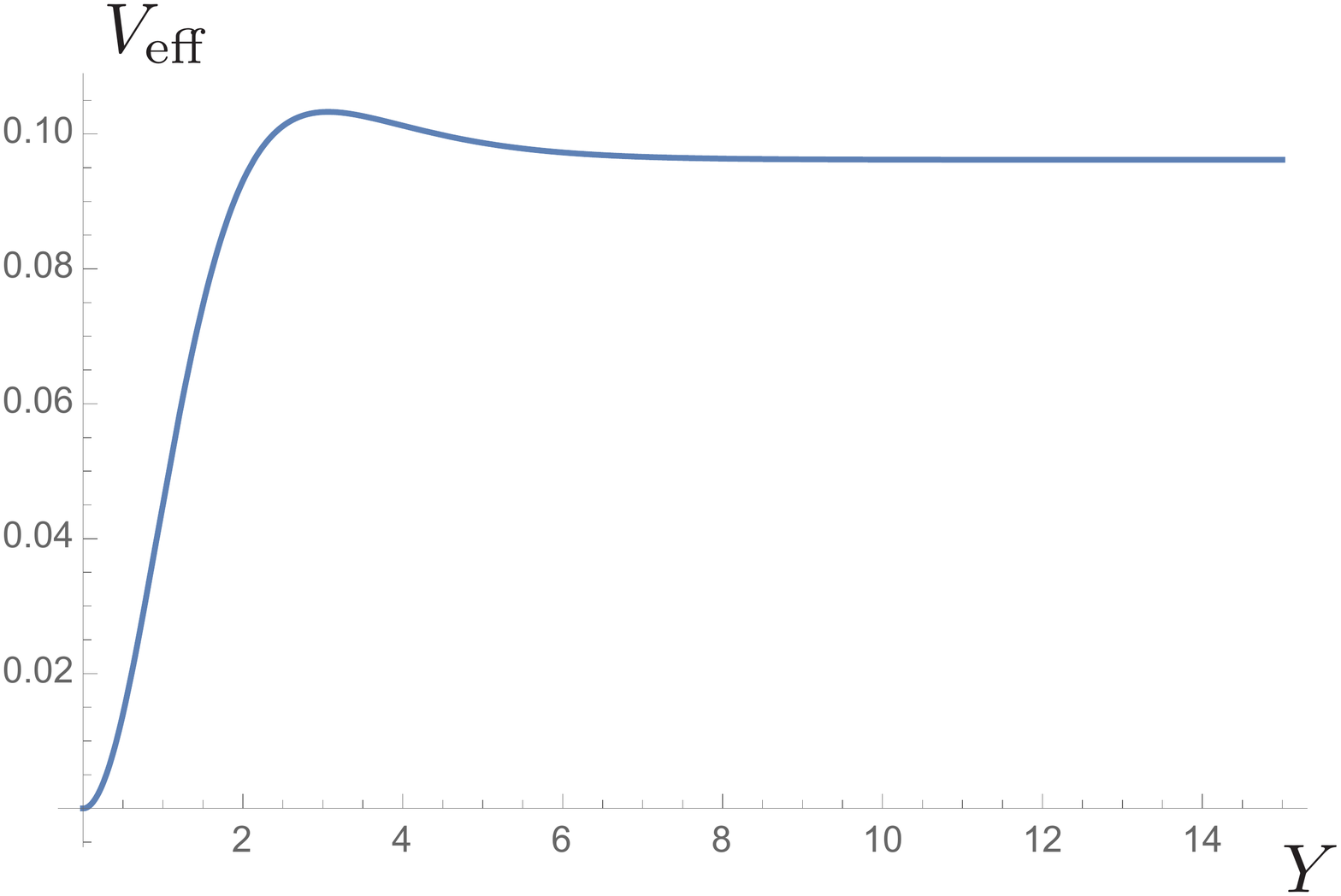} 
\caption{Interaction potentials for $(1,1)$ kink (left) and $(1,-1)$ kink (right) with $m=1$, $\beta=-1/10$, $X=5$.}
\label{fig:Veff}
\end{center}
\end{figure}

Figure\,\ref{fig:negative_(1,1)} shows the sine-Gordon solitons on domain walls  
with $\beta_{12}=\beta_{23}=\beta_{31}=-1/10$ and various values of $X$. 
We call these solitons ``$(1,1)$ kinks", 
since each phase degree of freedom has a single winding number. 
As shown in the figure with $X=5$, 
the two solitons tend to merge with each other for large $X$; 
i.e., there is an attractive force between them.  
The leading order interaction potential for large $X$ can be obtained by
substituting the two sine-Gordon kink configurations, 
\beq
\theta_{12} = 4 \arctan \exp \left[ \sqrt{\pi |\beta|}(y-Y) \right], \hs{10} 
\theta_{23} = 4 \arctan \exp \left[ \sqrt{\pi |\beta|}(y+Y) \right],
\eeq
into the domain-wall effective action, 
since this is a solution of the equation of motion in the large-$X$ limit. 
The interaction between $\theta_{12}$ and $\theta_{23}$ gives the leading order term 
\beq
V_{\rm int}(Y) = 2 \beta m X e^{-mX} \int dy \, \cos( \theta_{12} + \theta_{23} ) + {\it O}(e^{-mX}).
\eeq
Since the phases $\theta_{12}$ and $\theta_{23}$ tend to align with each other, 
the interaction potential is minimized when $Y=0$ 
and hence there is a attractive force for large $X$ 
(see the left panel of Fig.\,\ref{fig:Veff}). 
Ignoring $Y$-independent terms, we find that
\beq
V_{\rm int}(Y) = 16 m X e^{-mX} \sqrt{\frac{|\beta|}{\pi}} 
\left[ 1 - \frac{2 \sqrt{\pi |\beta|} Y}{\sinh\left( 2\sqrt{\pi |\beta|} Y \right)} \right] \coth^2 \left( \sqrt{\pi |\beta|} Y \right) + {\it O}(e^{-mX}).
\label{eq:Vint} 
\eeq

On the other hand, as $X$ becomes smaller,  
the two kinks start to depart from each other, 
as can be seen in the figure with $X=1,0$ and $-5$ 
in Fig.~\ref{fig:negative_(1,1)}.
Thus, the interaction becomes repulsive for small $X$.  
In the $X \rightarrow - \infty$ limit, 
this configuration is reduced to two sine-Gordon kinks in the single-wall effective action
and their interaction is known to be repulsive. 
Note that configurations in Fig.\,\ref{fig:negative_(1,1)} are quasistable,
implying that 
the distance between the kinks becomes larger as the fictitious time goes by. 

\begin{figure}[h]
\begin{center}
\includegraphics[width=22cm]{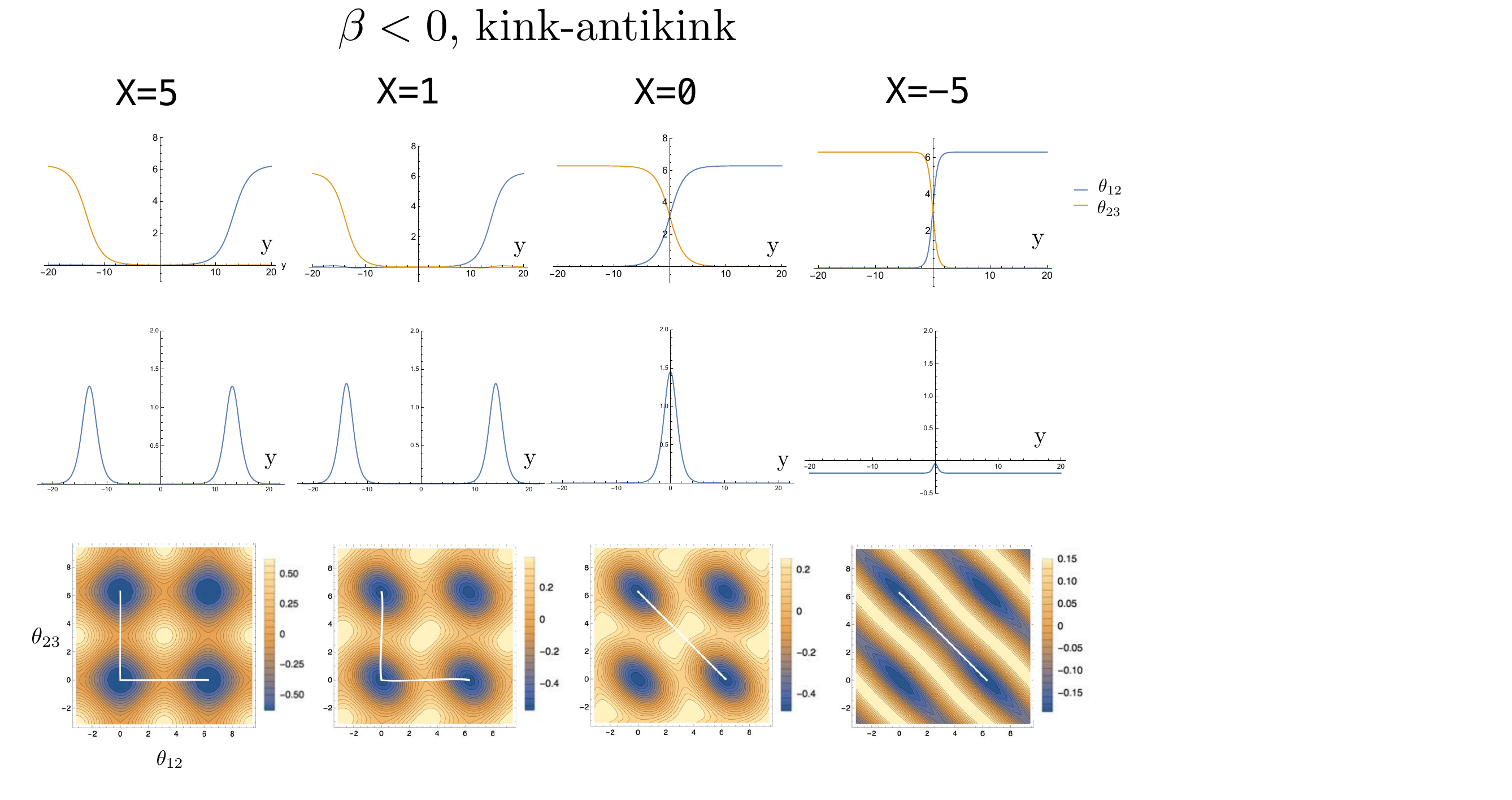}
\caption{$(1,-1)$ solitons for $X=5,1,0,-5$ with $\beta_{12}=\beta_{23}=\beta_{31}=-1/10$. 
The composition of the figure is the same as that in Fig.\ \ref{fig:negative_(1,1)}.
Note that the soliton profiles at $X=5,1$ are quasistable 
and there is a weak repulsive force between them.  
}
\label{fig5}
\end{center}
\end{figure}

\subsubsection{$(1,-1)$: the vortex-anti-vortex interaction}  
Figure\,\ref{fig5} shows configurations of ``$(1,-1)$-kink."  
In this case, the asymptotic interaction potential between the kink and antikink takes the form
\beq
V_{\rm int} = 16 m X e^{-mX} \sqrt{\frac{|\beta|}{\pi}} 
\left[ 1 + \frac{2 \sqrt{\pi |\beta|} Y}{\sinh \left( 2\sqrt{\pi |\beta|} Y \right)} \right] \tanh^2 \left( \sqrt{\pi |\beta|} Y \right) + {\it O}(e^{-mX}).
\label{eq:Vint(1,-1)} 
\eeq
As can be seen from Fig.\,\ref{fig:Veff}, 
the asymptotic interaction for large $X$ is repulsive for large $Y$ and attractive for small $Y$. 
Actually, for $X=5$, we have checked by numerical calculations 
that the interaction is repulsive for large $Y$ 
and attractive for small $Y$. The interaction changes its sign around $Y=3.06$. 
These results are consistent with the expectations from the potential of Eq.\,\eqref{eq:Vint(1,-1)}.  

As two domain walls approach each other, 
the repulsive force at large $Y$ changes to an attractive one 
suddenly at $X \simeq 0.144$, implying 
the attraction for all range of $Y$.

At $X=-5$, the potential is almost constant along $\theta_{12}+\theta_{23}={\rm const.}$ 
and there is almost no localized energy. 
This means that the kink on a domain wall and antikink 
on the other domain wall annihilate each other. 

\subsection{$\beta>0$: $\pi$-phases and frustration}
Next, we study the properties of sine-Gordon solitons  
for $\beta_{12}=\beta_{23}=\beta_{31} \equiv \beta >0$.

\subsubsection{Ground state}

In this case, the ground state structure changes depending on the 
distance of two walls. 

\begin{figure}[p]
\begin{center}
\hs{7}
\includegraphics[width=3.5cm]{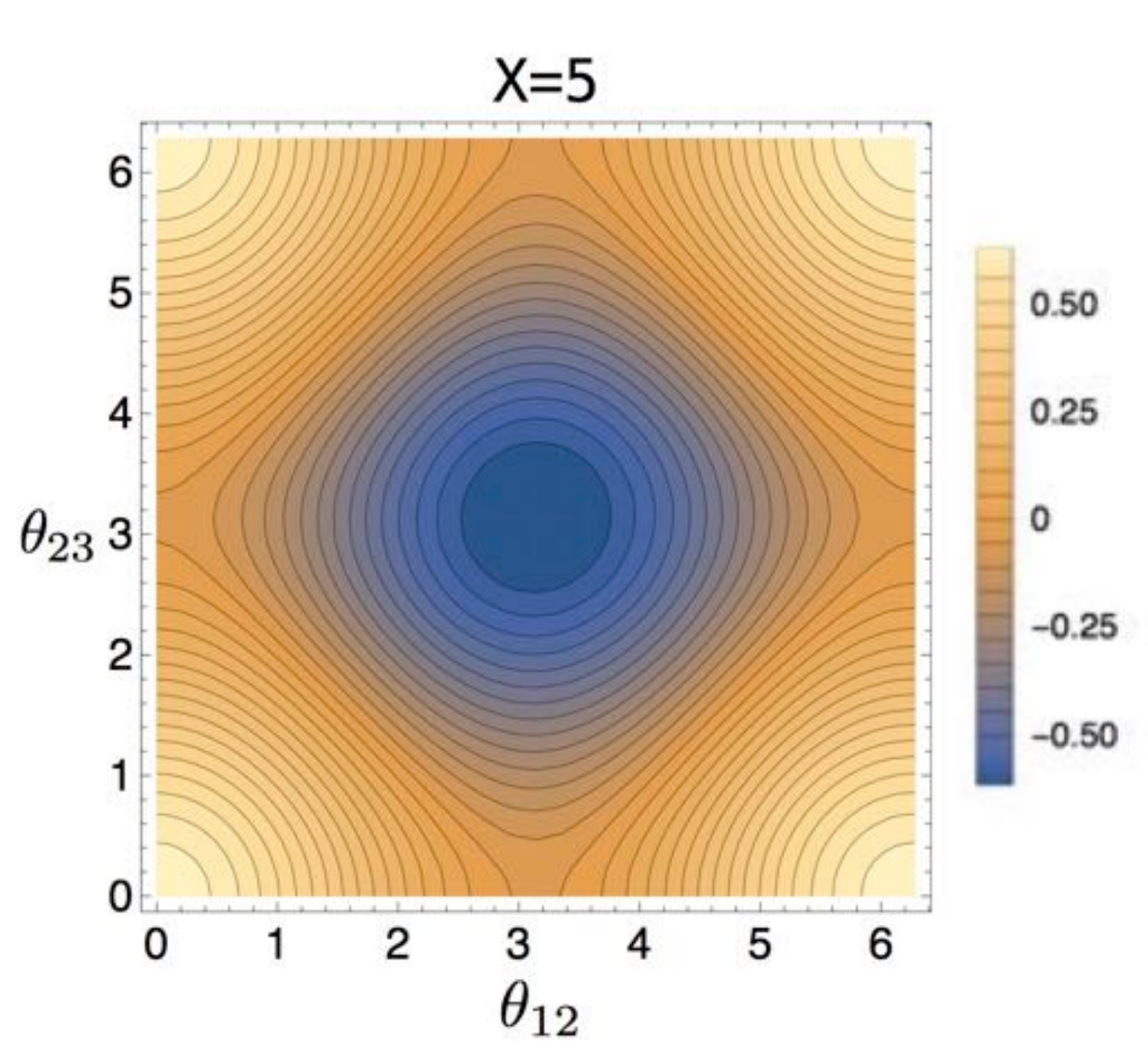}
\includegraphics[width=3.5cm]{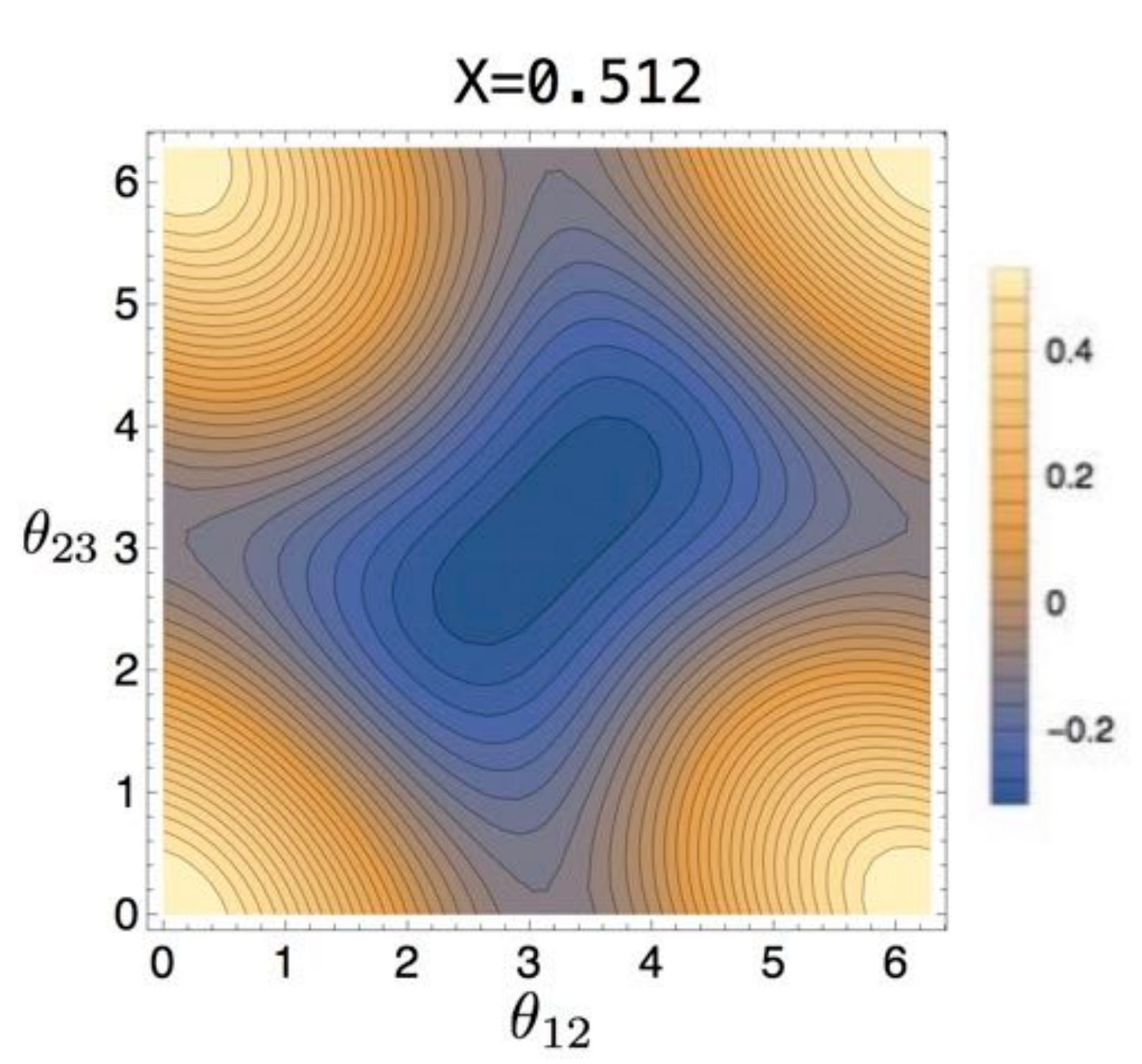}
\includegraphics[width=3.5cm]{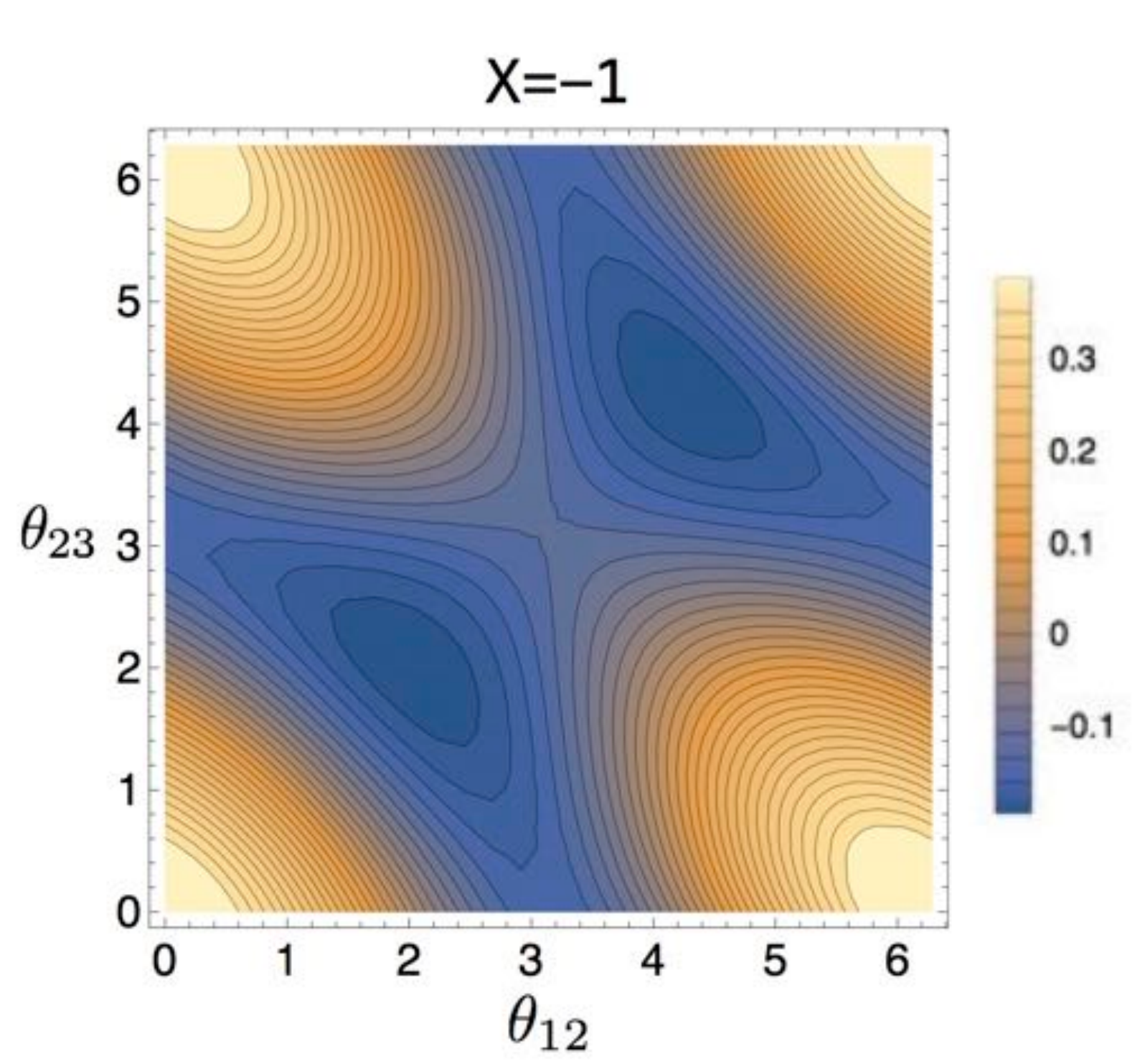}
\includegraphics[width=3.5cm]{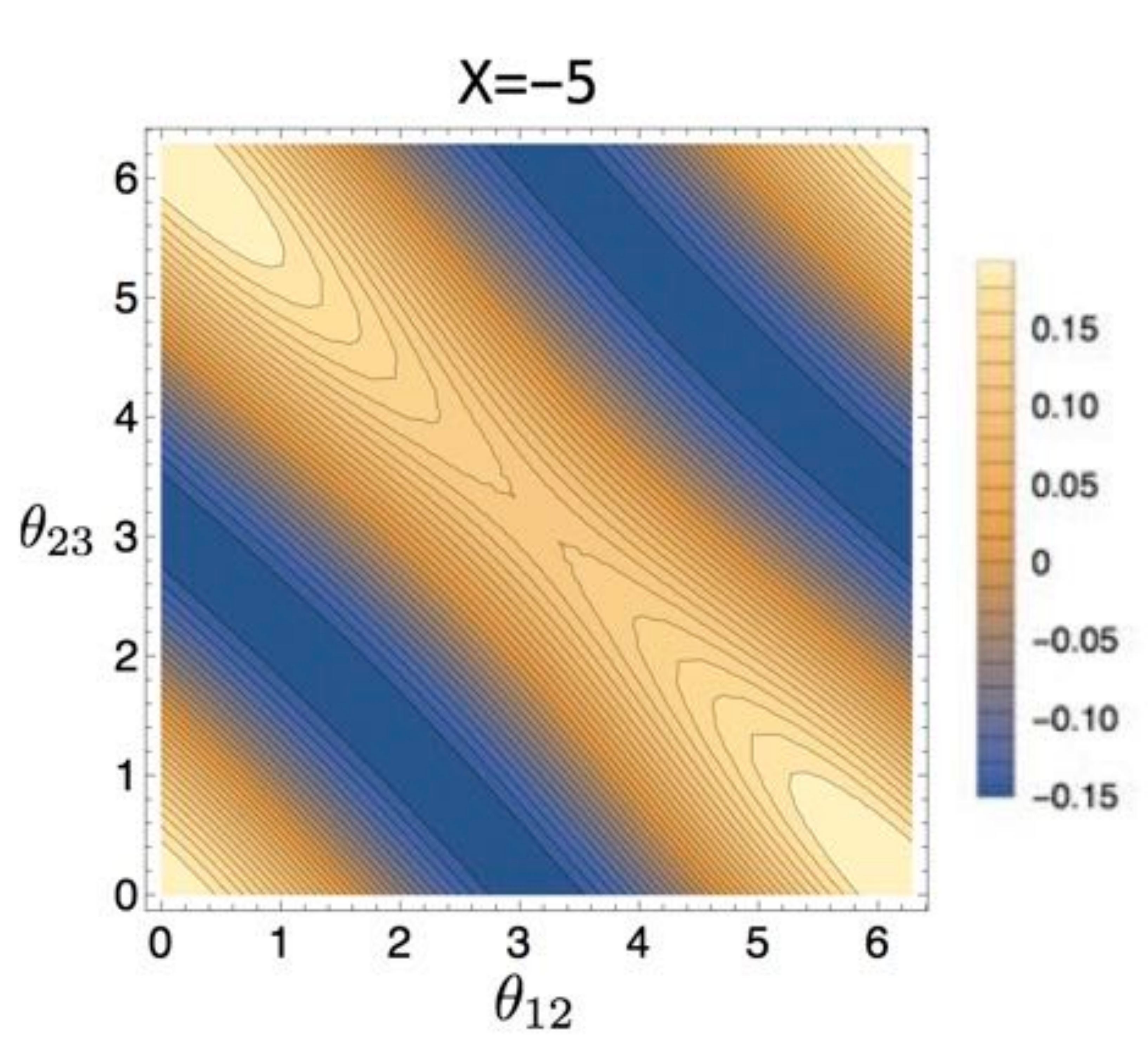}
\caption{The effective potential $V_{\rm eff}$ for $X=5, 0.512 ({\rm critical \ value}), -1$ and $-5$ 
with $\beta_{12}=\beta_{23}=\beta_{31}=1/10$.}
\label{fig:Veff_SSB}
\vs{20}
\includegraphics[width=17cm]{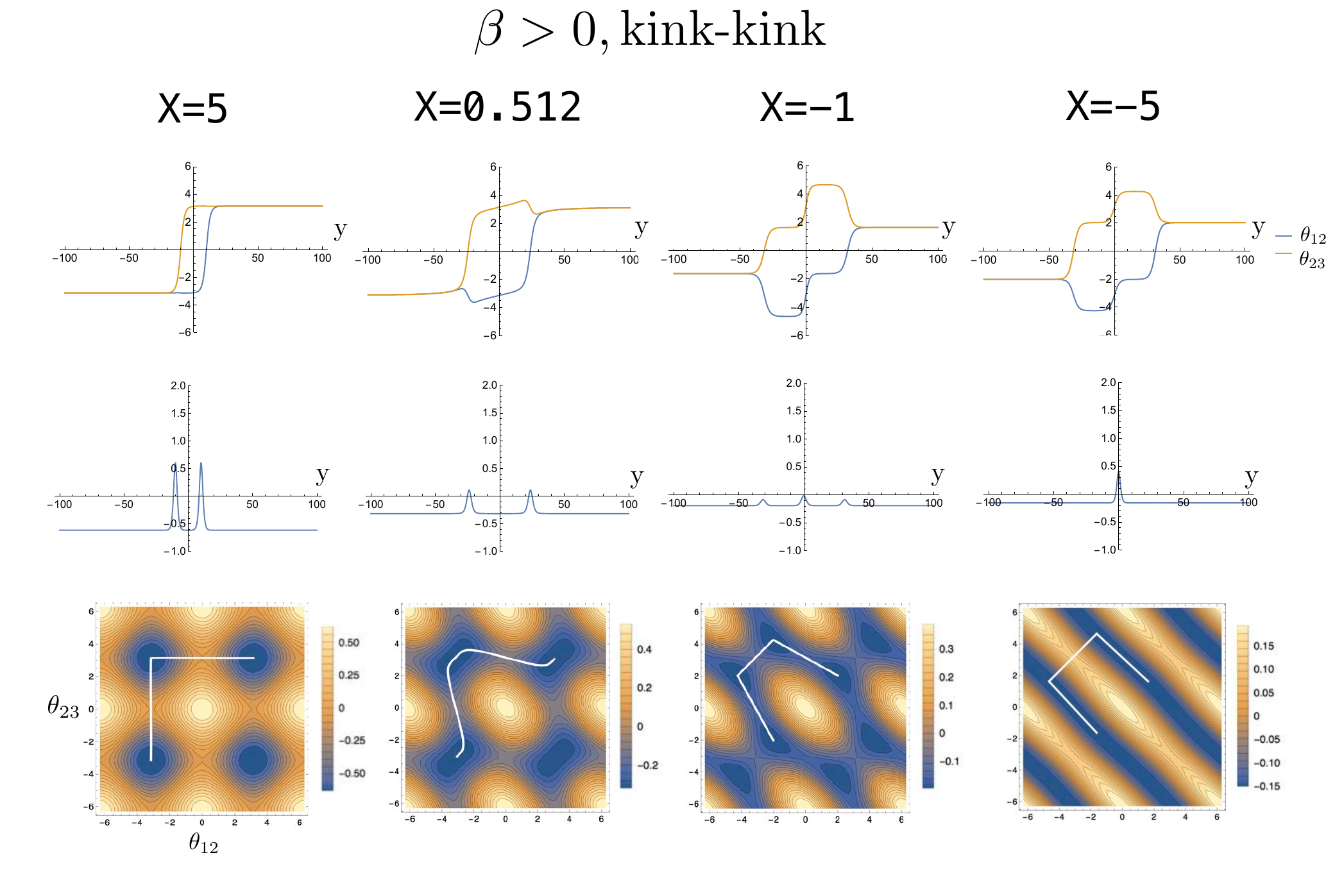}
\caption{$(1,1)$ solitons for $X = 5, 0.512$ (critical value)$, -1, -5$ 
with $\beta_{12}=\beta_{23}=\beta_{31}=1/10$. 
The upper panels show $y$ dependence of solitons, 
the middle panels are the energy densities, 
and the lower panels are the contour plots of $V_{\rm eff}$ and 
the soliton profiles in $\theta_{12}$-$\theta_{23}$ plane. 
The distance between two domain walls is denoted in the upper part of the figures. 
Note that $X=0.512$ is the critical value $X_c$, at which the two vacua emerge. }
\label{fig:positive_(1,1)}
\end{center}
\end{figure}

Figure\,\ref{fig:Veff_SSB} shows the effective potential $V_{\rm eff }$ at $\beta=1/10$. 
For large $X$, the term $F(X) ( \beta_{12} \cos\theta_{12} + \beta_{23} \cos\theta_{23})$ 
is dominant in $V_{\rm eff}$ and its minimum is located at $\theta_{12}=\theta_{23}=\pi$ (mod $2\pi$). 
On the other hand, as $X$ becomes smaller, 
the term $G(X) \beta_{31} \cos(\theta_{12}+\theta_{23})$, 
which has minima at $\theta_{12}+\theta_{23}=\pi$ (mod $2\pi$), becomes relevant. 
The two conditions, $\theta_{12}=\theta_{23}=\pi$ and $\theta_{12}+\theta_{23}=\pi$, 
cannot be satisfied simultaneously and hence there is a frustration for small $X$. 

We can easily see that $\theta_{12}=\theta_{23}=\pi$ is a stationary point of $V_{\rm eff}$:
\beq
d V_{\rm eff} \big|_{\theta_{12}=\theta_{23}=\pi} = 0.
\eeq
At this point, the charge conjugation symmetry $\phi_a \rightarrow \bar \phi_a$ is preserved. 
The Hessian (the determinant of the second derivatives) of $V_{\rm eff}$ around the stationary point
$\theta_{12}=\theta_{23}=\pi$ is given by
\bal
H = \frac{\pi\beta^2}{1+2e^{-mX}} \big[ 1-  f(X) \big], \hs{10} 
f(X) \equiv \frac{4}{\pi} e^{-mX}\frac{\cosh^{-1}\frac{e^{mX}}{2}}{\sqrt{1-2e^{-mX}}}.
\eal
Here $f(X)$ is a monotonically decreasing function such that $f(X \rightarrow - \infty) = + \infty$ and $f(X \rightarrow \infty) = 0$
and hence there is a critical value $X_c$ at which $H$ changes its sign. 
Since $H>0$ for large $X$, 
the point $\theta_{12}=\theta_{23}=\pi$ is a stable minimum
when two domain walls are well separated. 
On the other hand, for $X<X_c$, 
the minimum splits into a pair of points 
which are exchanged by the charge conjugation. 
Thus, $X_c$ is the critical value at which 
the charge conjugation symmetry is spontaneously broken. 
For $\beta_{12}=\beta_{23}=\beta_{31}$, 
the critical value is 
\begin{equation}
m X_c \simeq 0.512. \label{eq:Xc}
\end{equation}

\subsubsection{$(1,1)$: the vortex-vortex interaction}  

Figure\,\ref{fig:positive_(1,1)} shows the configurations of $(1,1)$ solitons with $X=5,\,0.512,\,-1,\,-5$. 
At $X=5$, the two solitons weakly repel each other. 
The asymptotic interaction potential between them takes the same form as Eq.\,\eqref{eq:Vint}
with the opposite sign. 
At $X=X_c=0.512$, the minimum of the potential splits into the pair of vacua  
and there emerges another kink connecting them. 
For small $X$, solitons connect the two vacua 
$(\theta_{12},\theta_{23})=(-\pi/2,-\pi/2)$, $(\pi/2,\pi/2)$
as shown in the right figures in Fig.\,\ref{fig:positive_(1,1)}. 
Although it appears that there are three kinks (see $X=-1$ case in Fig.\ref{fig:positive_(1,1)}), 
two of them have very small energy 
since $\theta_{12}-\theta_{23}$ becomes unphysical as $X \rightarrow -\infty$.  
Thus, only one kink is left in the small-$X$ limit. 

Distribution of magnetic fluxes in the Josephson junction 
gives the important information of Josephson vortices, 
and is one of the main subjects of studies of the Josephson effect 
\cite{Liu2011,Berdiyorov1,Berdiyorov2,Berdiyorov3,Berdiyorov4,Berdiyorov5}. 
For the vortex-vortex interaction in the frustrated case, a remarkable consequence of the change of vacuum structure 
can be seen in the magnetic flux: 
\bal
\frac{1}{2\pi} \int d x d y \, F_{xy} ~=~ \frac{1}{2\pi} \oint dx^\mu \frac{i}{2}(\bar\phi^a\partial_\mu\phi^a 
-(\partial_\mu \bar\phi^a)\phi^a) 
~=~ \frac{1}{2\pi} \left[\theta_{\rm 12}+\theta_{\rm 23}\right]^{y=+\infty}_{y=-\infty}.
\eal

\begin{figure}[t]
\begin{center}
\includegraphics[width=17cm]{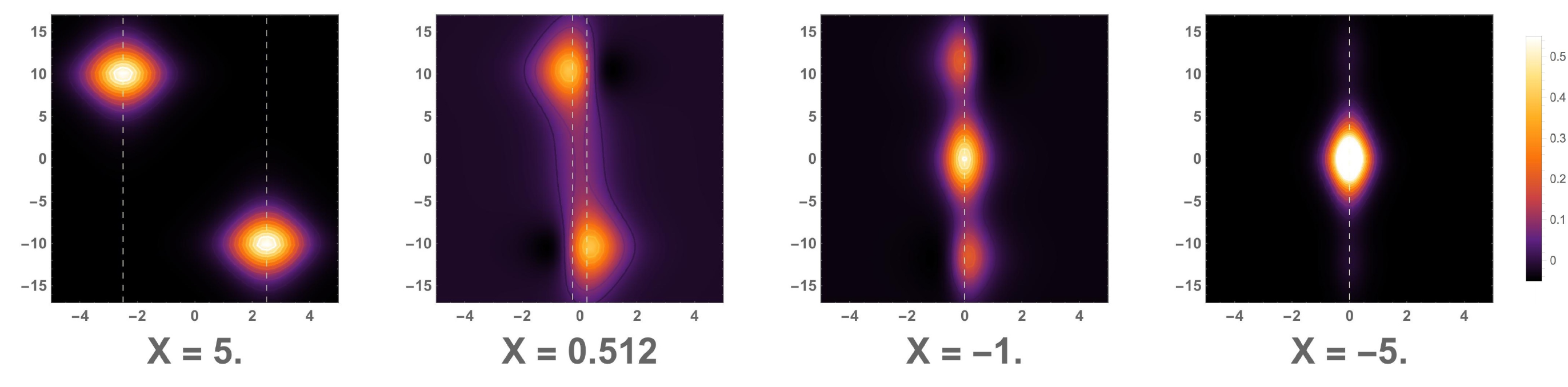} \hs{10}
\caption{The magnetic flux distribution in the $x$-$y$ plane. 
The dashed vertical lines show the positions of the domain walls.}
\label{fig:flux_distribution}
\end{center}
\end{figure}
\begin{figure}[t]
\begin{center}
\includegraphics[width=7cm]{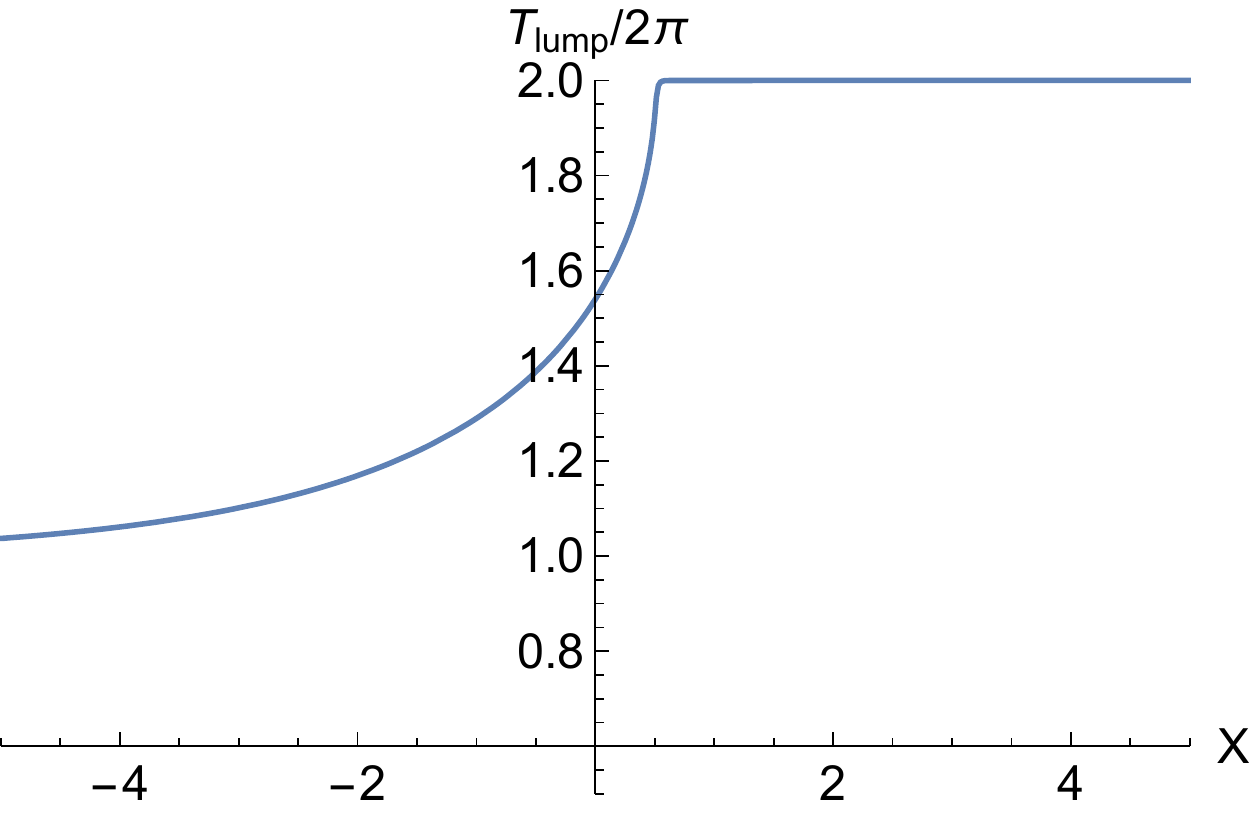}
\caption{The magnetic flux of solitons corresponding to Fig.\ref{fig:positive_(1,1)}.}
\label{fig:flux}
\end{center}
\hspace{0.9cm}
\end{figure}
Figures.\,\ref{fig:flux_distribution} and \ref{fig:flux} show 
the magnetic flux distribution and the total flux of the solitons in Fig.\,\ref{fig:positive_(1,1)}, 
respectively. 
For $X>X_c=0.512$, there are two units of magnetic flux. 
On the other hand, for $X<X_c$, 
the flux begins to decrease due to the emergence of two vacua: 
each soliton becomes a ``fractional soliton,"  
which connects a pair of inequivalent vacua. 
The configuration in the small-$X$ limit has one unit of magnetic flux  
since the solitons connect $(\theta_{12},\theta_{23})=(-\pi/2,-\pi/2)$ and $(\pi/2,\pi/2)$.


\subsubsection{$(1,-1)$: the vortex-anti-vortex interaction}  

Figure\,\ref{fig:positive_(1,-1)} shows the configurations of $(1,-1)$ solitons. 
At $X=5$, both the kink and antikink are located around $y \approx 0$. 
The asymptotic interaction potential between them takes the same form as 
Eq.\,\eqref{eq:Vint(1,-1)} with the opposite sign. 
Since the minimum is located at $Y \not = 0$, 
the kink and anti-kink keep a small distance between them. 
As $X$ becomes smaller, 
the solitons change their form and their masses gradually decrease due to the change of the potential. 
Finally, no localized energy is left at $X=-5$ 
since only the relative phase $\theta_{12}-\theta_{23}$ has kinks and 
it becomes unphysical in the small-$X$ limit. 
Note that flux is always zero in this case.

\subsubsection{$(2,2)$: the vortex-vortex interaction}  

Finally, we show the $(2,2)$ solitons in Fig.\ref{fig:wall_profile0} and \ref{fig:wall_profile1}. 
We can see similar behaviors to those in the $(1,1)$ case shown in Fig.\ref{fig:positive_(1,1)}. 
As shown in Fig.\ref{fig:wall_profile1}, 
the magnetic flux, which is initially four at $X=5$, begins to decrease at $X=X_c$ in Eq.~(\ref{eq:Xc}), 
and then reduces to three in the small $X$ limit. 

\section{Summary and discussions}
\label{sec5}

We have proposed the $\C P^{N-1}$ model 
as a model to describe an $N$-layered Josephson junction.
To illustrate use, 
we have studied 
dynamics of Josephson vortices 
by studying 
the sine-Gordon solitons on multiple domain walls. 
For $N=3$, we have investigated the two cases in which the charge conjugation symmetry is preserved. 
When the coupling constants $\beta_{ab}$ are all positive, 
the vacuum structure on the domain walls is independent of their relative distance, 
whereas the structure changes at a critical distance $X=X_c$ when $\beta_{ab}$ are all negative.
 
In the former case, the interaction between 
Josephson vortices on different domain walls changes 
with the distance $X$ between two domain walls. 
For $(1,1)$ solitons (kink-kink configuration), 
the interaction is attractive at large $X$ and 
repulsive at small $X$.  
In the case of the $(1,-1)$ solitons (kink-antikink configuration), 
the interaction is repulsive at large $Y > 3.06$ and attractive at small 
$Y<3.06$ for $X>0.144$, 
while it becomes attractive for all ranges of $Y$ 
for $X<0.144$. 
 
In the latter case, the properties of the Josephson vortices change  
depending on the distance between the domain walls. 
There is a critical value $X=X_c$ at which the charge conjugation symmetry is spontaneously broken on the domain walls.   
For $X>X_c$, the total magnetic flux is constant,  
whereas for $X<X_c$, the flux gradually decreases as $X$ becomes smaller 
and hence there emerge fractional sine-Gordon solitons. 

Here we comment on the related studies with our 
frustrated multiband superconductors.  

In Ref.\,\cite{Leggett1966}, 
the system of two-band superconductors is discussed and 
the collective excitation with respect to the fluctuations of 
the relative phase of two condensates is found,  
which is a kind of Josephson effect. 
The excitation, the Leggett mode, is actually observed in experiments on 
Mg-B$_2$ \cite{Blumberg2007,Nagamatsu2001}. 
Theoretically, the excitation with fractional flux quanta is discussed 
in several works \cite{Sigrist1999,Tanaka2001,Babaev2002,Gurevich2003,
Bluhm2006,Vakaryuk2012}. 
In the stream of the studies, 
the system with three or more condensations and frustration between them 
has recently been given attention. 
In Ref.~\cite{Stanevetal2010}, 
the system was studied in which  
three superconductor bands are connected via repulsive pair-scattering terms, 
where a time-reversal-symmetry breaking (TRSB) state emerges. 
The Ginzburg-Landau theory is derived from the multiband BCS Hamiltonian 
in the general case in Ref.~\cite{Orlovaetal2013}. 
In Refs.~\cite{Huangetal2014,Huangetal2015,Huangetal2015}, 
Josephson junctions between chiral and regular superconductors were considered: 
the asymmetric critical currents, subharmonic Shapiro steps,  
symmetric Fraunhofer patterns \cite{Huangetal2014}, 
and the fractional flux and its plateau in magnetization curve 
\cite{Huangetal2015,Huangetal2016} are studied by using 
Bogoliubov--de Gennes and the time-dependent Ginzburg-Landau equation.  
Also, in Ref.\cite{Takahashietal2014}, the phase diagram of the system was investigated 
in the $H$-$T$ plane. 


Here we address several discussions.


In this system, the plasma oscillations occur.  
Due to the change of vacuum structure, 
the properties of plasma oscillations, such as dispersion relations, 
vary with the positions of domain walls. 
This is a peculiar property for the multilayered Josephson junctions. 
The analysis will be reported elsewhere. 

It is curious as to whether there is a real system described by the model. 
The system constructed of three superconductors and 
two thin insulators in between may be described by the model. 
If the strength of the couplings can be changed  
and the distance between the two insulators can be controlled, 
we can see the change of the interaction between the solitons, 
and the emergence of the fractional sine-Gordon solitons. 

We consider another possible experimental setup  
than the normal layers, 
which is pictorially shown in Fig.~\ref{appr_setup}.  
Superconductors 1 (sc1), 2 (sc2), and 3 (sc3) are divided by thin insulators (black lines). 
The sc2 has the form of an acute-angled triangle. 
In this setup, the pairwise coupling of sc1 and sc3 is dominant in the upper part. 
On the other hand,  
the couplings between sc1 and sc2, and sc2 and sc3, become dominant in the 
lower part. 
There occurs frustration around the node of the insulators, 
and we could see the fractional vortices on the thin insulators. 
The distance of sc1 and sc3 is spatially and moderately dependent 
on the position of the vertical direction of Fig.~\ref{appr_setup}. 
We may realize the situation that we want somewhere in the vertical direction. 
If we can make the setup artificially or accidentally in experiment, 
and make many vortices on the insulators especially around the node, 
we may observe a fractional vortex by manipulating a vortex 
using the scanning tunneling microscope and by placing it on the node.  

An appropriate setup might be also realized in 
Bose-Einstein condensates (BECs) of ultracold-atomic gases. 
Mixture of two or more condensates 
of hyperfine states of a single atom
provide multicomponent BECs. 
When they are repulsive 
a phase separation occurs to form domain walls.
We can introduce Rabi oscillations 
to provide Josephson couplings. 
In this case, in principle, one might consider 
both  unfrustrated \cite{Eto:2012rc} 
and frustrated \cite{OKM2016} cases. 

In this paper, we have regarded the domain walls as infinitely heavy and analyze 
the sine-Gordon kinks by fixing the positions of 
the domain walls.  
Without such an assumption, the domain walls 
can move giving flexible Josephson junctions.
The analysis of full dynamics of the system, i.e., 
the time and space dependence of domain walls and the sine-Gordon solitons on the walls, should be interesting, 
as in Ref.~\cite{Jennings:2013aea} for the two component case.

The model admits a $Y$ junction of domain walls 
which meet at a junction point
\cite{Eto:2005cp}, 
if we introduce complex masses $m$ for $\phi_a$.
More generally, the model admits a network of 
junctions. 
The effective action of such a network was obtained in 
Ref.~\cite{Eto:2006bb}.
This can be applied to a $Y$-shaped insulator
of Josephson junctions of three superconductors 
if we introduce Josephson interactions.
Introducing Josephson interaction to this case is an interesting problem.

In this paper, we applied magnetic field in parallel with 
insulators 
so that vortices are absorbed along the insulators 
to become Josephson vortices.
If we apply magnetic field orthogonal to
the insulators,  magnetic vortices 
end up with the insulators, where two magnetic vortices 
in neighboring superconductors 
are connected by pancake vortices 
\cite{Blatter:1994}.
The same configurations without the Josephson interaction 
is a D-brane soliton \cite{Gauntlett:2000de,Isozumi:2004vg}.
In particular, the most general analytic solutions 
in the ${\mathbb C}P^{N-1}$ model 
(relevant for layered Josephson junctions)
was obtained in Ref.~\cite{Isozumi:2004vg}.
The effective action and dynamics of such a system 
were studied in Ref.~\cite{Eto:2008mf} 
without the Josephson interaction.
Introducing the  Josephson interactions 
in this system should be interesting for the 
study of pancake vortices in field theory.

If we consider a quadratic Josephson term 
$|\bar \phi_a \phi_b|^2$
instead of the linear Josephson term 
$\bar \phi_a \phi_b$ considered in this paper, 
the system can be made 
supersymmetric by appropriately adding fermions
as was shown for the ${\mathbb C}P^1$ case \cite{Auzzi:2006ju}.
In this case, the minimum Josephson vortices
carry half fluxes, 
and the total configurations are 1/4 BPS 
preserving a quarter of supersymmetry.
The situation should be the same for the case of 
the multilayered Josephson junction studied in this paper.

Domain-wall solutions in
non-Abelian gauge theory were constructed 
in Refs.~\cite{Isozumi:2004jc,Eto:2006pg}.
A non-Abelian generalization of Josephson junctions 
was proposed in Refs.~\cite{Nitta:2015mma}
in which a junction of two non-Abelian $U(N)$ (color) superconductors 
was discussed.
The low-energy effective action of the 
non-Abelian domain wall (insulator) 
can be described by a $U(N)$ chiral Lagrangian 
\cite{Eto:2005cc}  
with a pion mass term (non-Abelian sine-Gordon model)
\cite{Nitta:2014rxa},  
admitting a non-Abelian sine-Gordon soliton 
\cite{Nitta:2014rxa,Yanagisawa:2016uia} 
which corresponds to 
a non-Abelian Josephson vortex \cite{Nitta:2015mma}. 
A multilayered non-Abelian Josephson junction is 
one of the possible future directions. 

\section*{Acknowledgments}
We would like to thank Zhao Huang for useful comments 
and discussions and Naoki Yamamoto for a discussion at the early stage of this work. 
This work is supported by the Ministry of Education, Culture, Sports, Science, and Technology (MEXT) Supported Program for the Strategic Research Foundation at Private Universities ``Topological Science'' ({Grant No.~S1511006}). 
The work of M.N. is also supported in part 
by a Grant-in-Aid for Scientific Research on Innovative Areas
``Topological Materials Science"
(KAKENHI Grant No. 15H05855) and 
``Nuclear Matter in Neutron Stars Investigated by Experiments and
Astronomical Observations"
(KAKENHI Grant No. 15H00841) 
from the MEXT of Japan 
and by a Japan Society for the 
Promotion of Science (JSPS) 
Grant-in-Aid for Scientific Research
(KAKENHI Grant No. 16H03984). 
H.I. was supported by the RSF grant 15-12-20008.

\begin{figure}
\includegraphics[width=18cm]{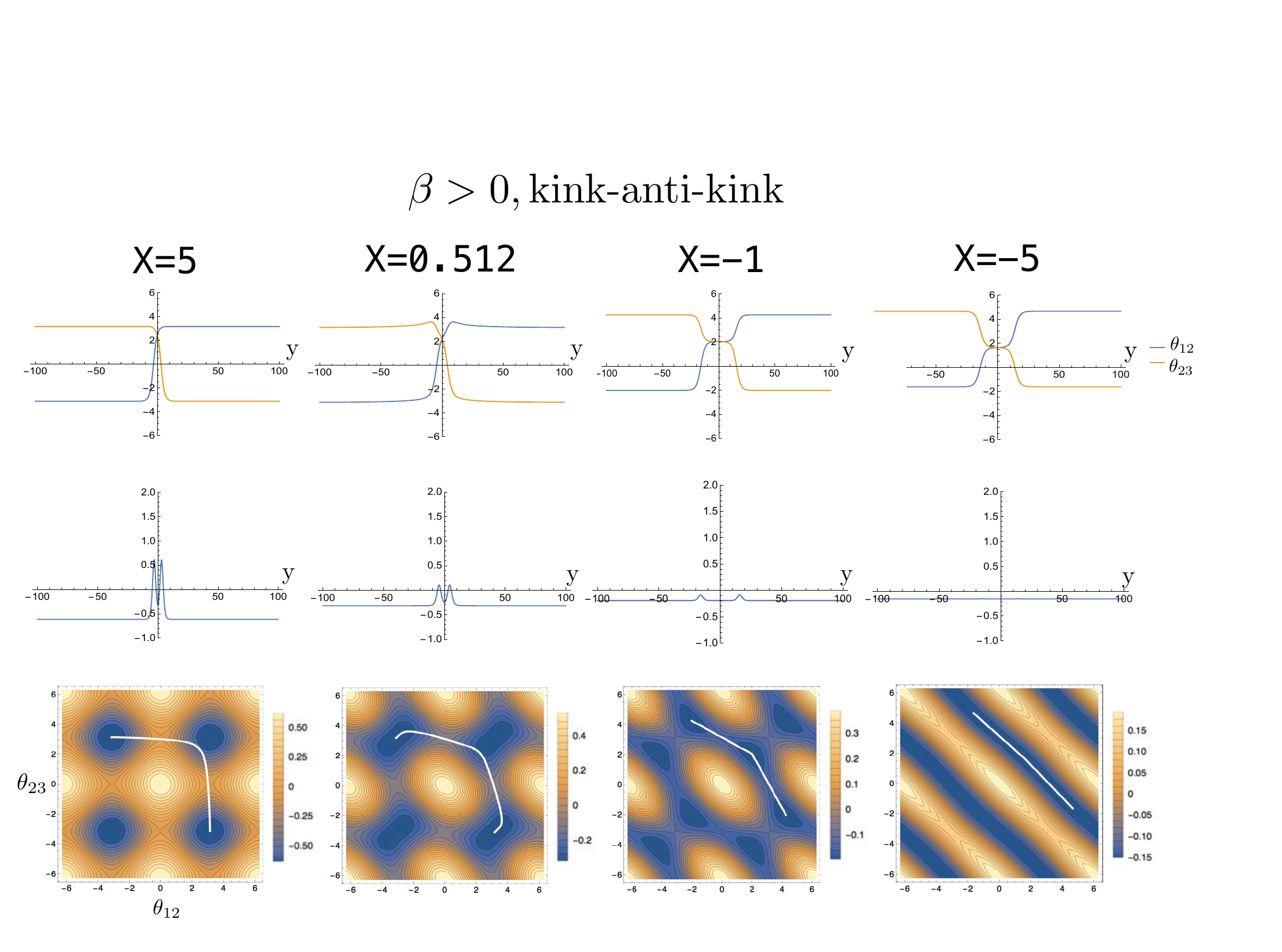}
\caption{One-kink/one-antikink solitons for $X = 5, 0.512$(critical value)$, -1, -5$ 
with $\beta_{12}=\beta_{23}=\beta_{31}=1/10$. 
The composition of the figure is the same as that in Fig.~\ref{fig:negative_(1,1)}. 
Note that the solitons are quasi-stable for $X=-5$.}
\label{fig:positive_(1,-1)}
\end{figure}

\begin{figure}
\includegraphics[width=17cm]{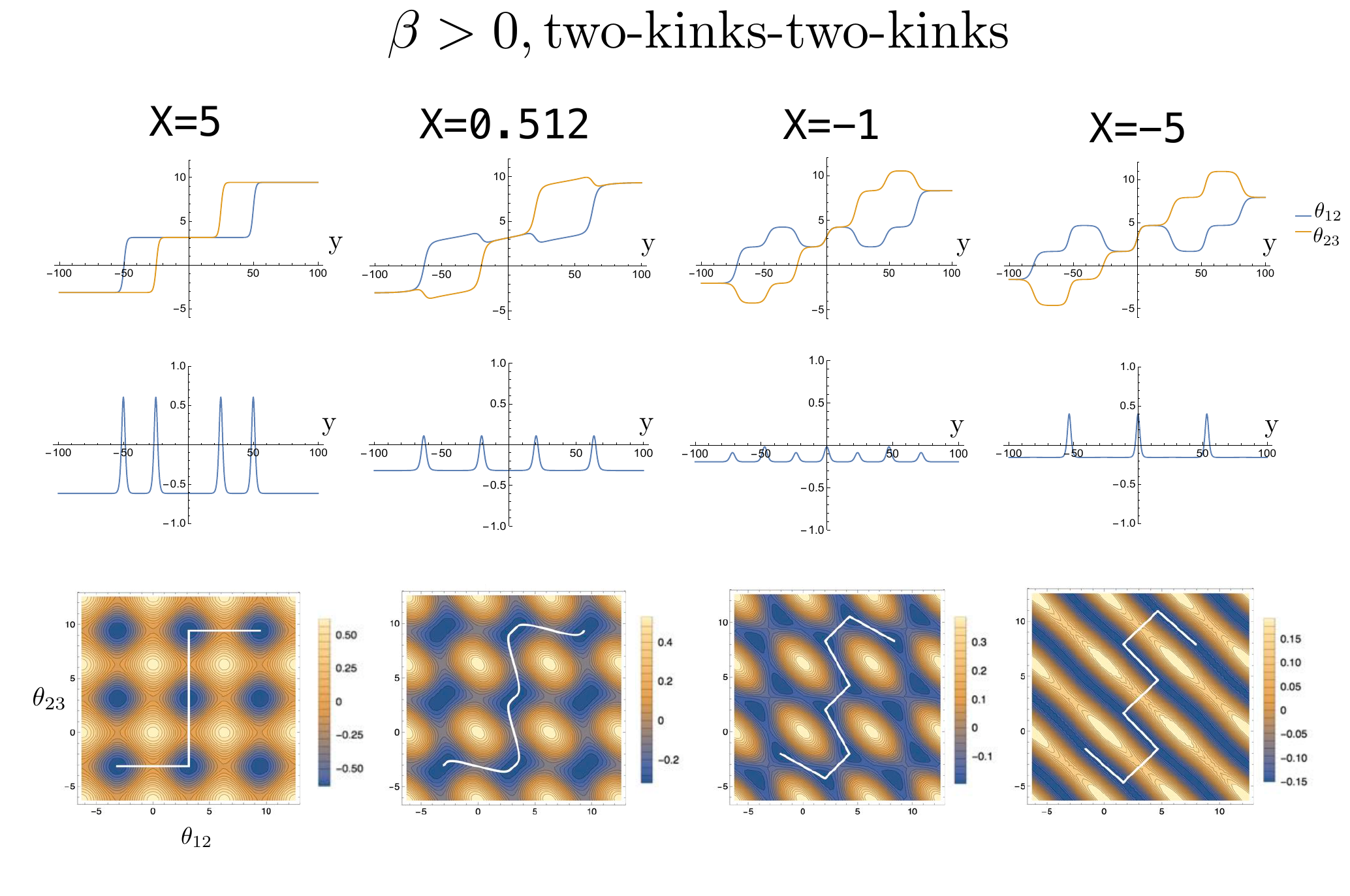}
\caption{
Two-kinks/Two-kinks solitons for $X = 5, 0.512$ (critical value)$, -1, -5$ with $\beta_{12}=\beta_{23}=\beta_{31}=1/10$. 
The composition of the figure is the same as that in Fig.~\ref{fig:negative_(1,1)}. 
Note that the solitons are quasistable.}
\label{fig:wall_profile0}
\hspace{0.9cm}
\end{figure}

\begin{figure}
\begin{center}
\includegraphics[width=7cm]{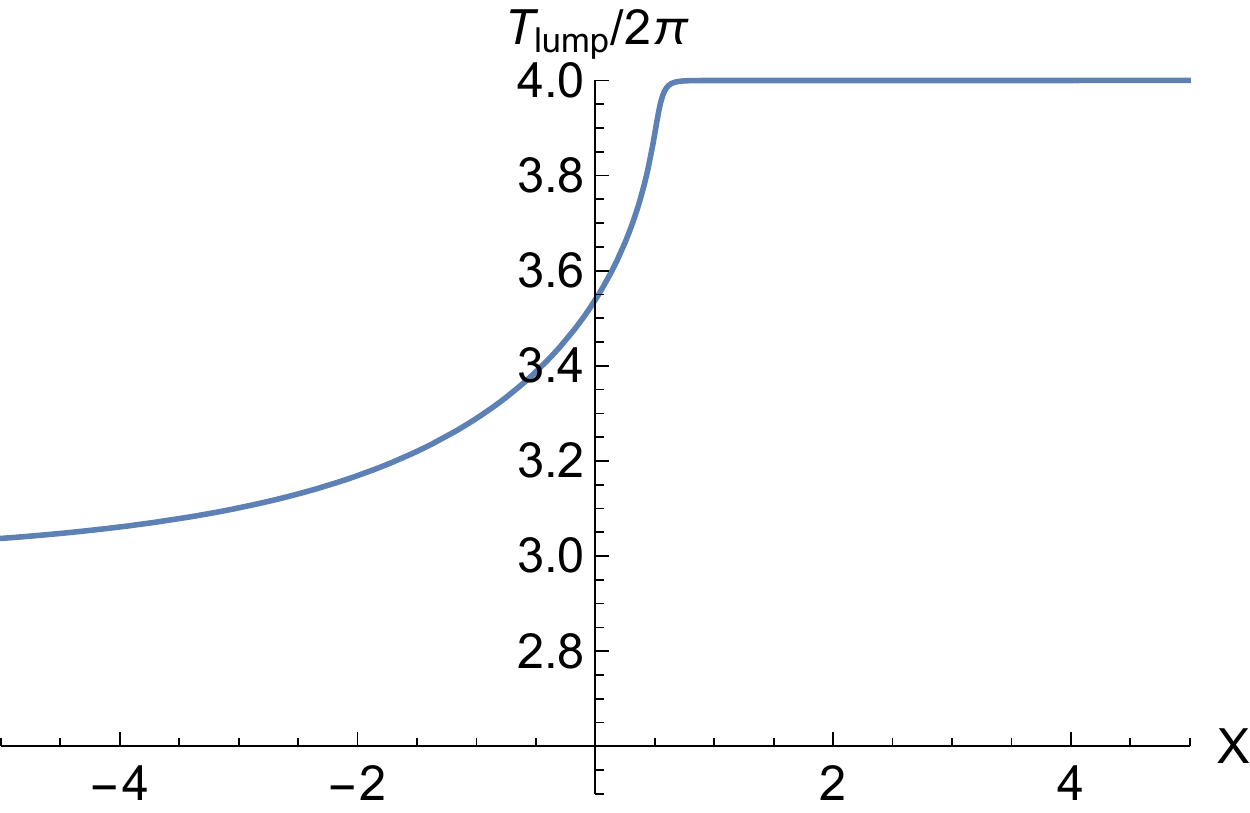}
\caption{The flux of solitons corresponding to Fig.\ref{fig:wall_profile0}. }
\label{fig:wall_profile1}
\end{center}
\hspace{0.9cm}
\end{figure}

\begin{figure}[t]
\begin{center}m
\includegraphics[width=4cm]{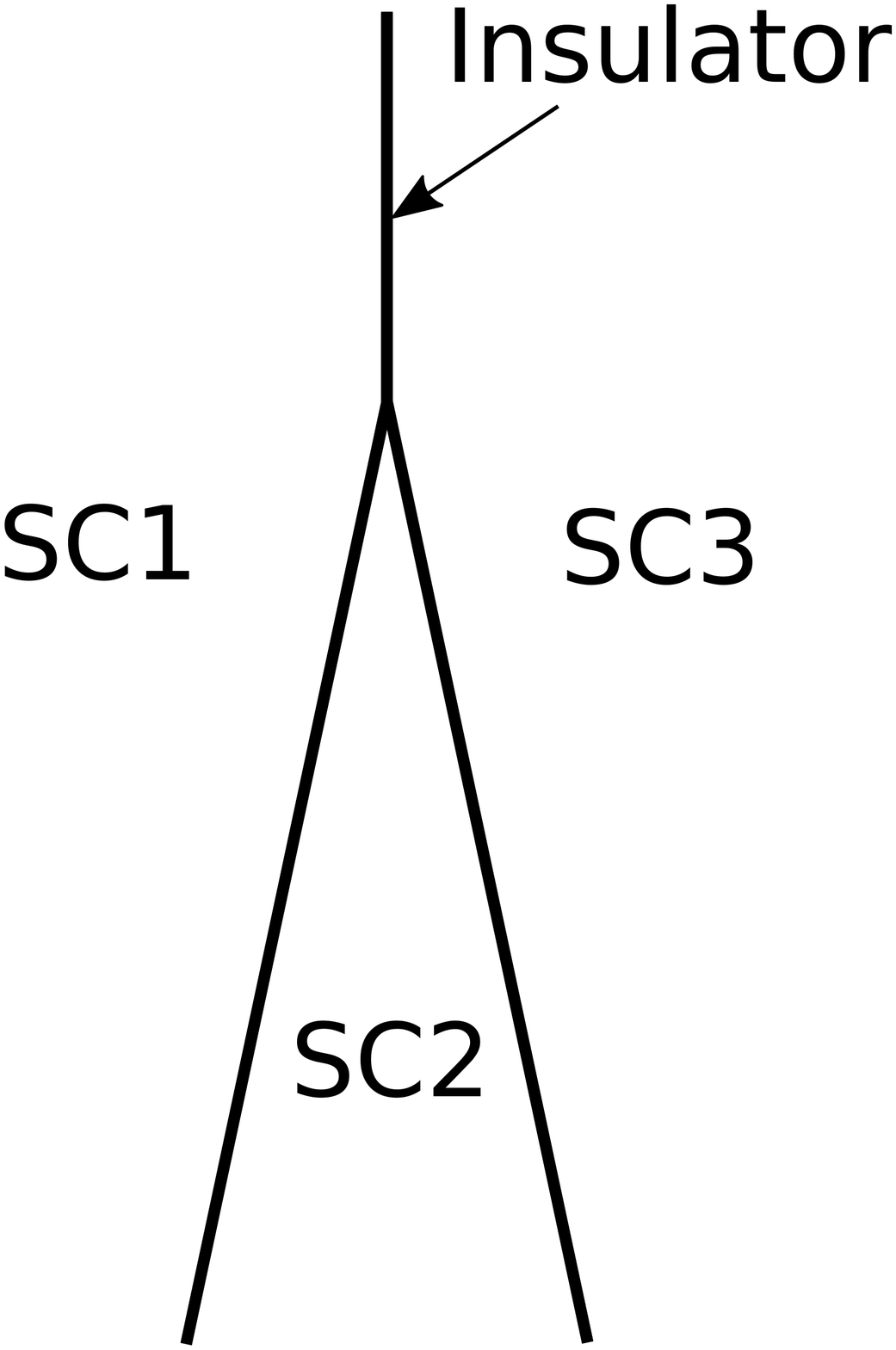}
\caption{A possible experimental setup for our study}
\label{appr_setup}
\end{center}
\end{figure}

\end{document}